\documentclass[prd,nofootinbib,showpacs,twocolumn]{revtex4-1}
\usepackage{color}
\usepackage{bm}
\usepackage{graphicx}%caption,subcaption
\usepackage{amsmath}
\usepackage{amssymb}
\usepackage{enumitem}
\usepackage{hyperref}
\usepackage{array}
\usepackage[english]{babel}
\usepackage{tensor}
\usepackage{comment}
\usepackage[normalem]{ulem}
\usepackage[dvipsnames]{xcolor}
\allowdisplaybreaks
\usepackage[utf8]{inputenc}
\usepackage{graphicx}  
\usepackage{txfonts}  
\usepackage{epstopdf}
\usepackage{color}

\def\be{\begin{equation}}
\def\ee{\end{equation}}
\def\ba{\begin{eqnarray}}
\def\ea{\end{eqnarray}}
\def\l{\left}
\def\r{\right}
\def\f{\frac}

\newcommand{\gsim}{\raisebox{-0.13cm}{~\shortstack{$>$ \\[-0.07cm]
      $\sim$}}~}

\usepackage{tikz}

\def\hub{{\mathcal H}}

\definecolor{orange}{rgb}{1,0.5,0}
\definecolor{darkorange}{rgb}{0.69,0.33,0.13}
\definecolor{darkgreen}{rgb}{0.05,0.5,0.06}

\newcommand\lcdm{$\Lambda$CDM }

\newcommand{\beq}{\begin{equation}}
\newcommand{\eeq}{\end{equation}}
\newcommand{\beqa}{\begin{eqnarray}}
\newcommand{\eeqa}{\end{eqnarray}}

\def\eftcamb{\texttt{EFTCAMB} }

\begin{document}

\title{Phenomenology of the generalized cubic covariant Galileon model and cosmological bounds}

\author{Noemi Frusciante$^1$, Simone Peirone$^2$, Lu\'is Atayde$^1$, Antonio De Felice$^3$}

\affiliation{\smallskip
$^1$Instituto de Astrof\'isica e Ci\^encias do Espa\c{c}o, 
Faculdade de Ci\^encias da Universidade de Lisboa,  Campo Grande, PT1749-016 Lisboa, Portugal
\smallskip\\
$^2$ Institute Lorentz, Leiden University, PO Box 9506, Leiden 2300 RA, The Netherlands
\smallskip\\
$^3$ Center for Gravitational Physics, Yukawa Institute for Theoretical Physics, Kyoto University, 606-8502, Kyoto, Japan}

\begin{abstract}

We investigate the generalized cubic covariant Galileon model, a kinetically driven dark energy model within the  Horndeski class of  theories. 
The model extends the cubic covariant Galileon by including power laws of the field derivatives in the K-essence and cubic terms which still allow for tracker solutions.  We study the shape of the viable parameter space by enforcing stability conditions which include the absence of ghost, gradient and tachyon instabilities and the avoidance of strong coupling at early time. 
 We study here the relevant effects of the modifications induced by the model on some cosmological observables such as the cosmic microwave background (CMB), the lensing potential auto-correlation and the matter power spectrum.  
For this goal, we perform parameter estimation using data of CMB temperature and polarization, baryonic acoustic oscillations (BAO), redshift-space distortions (RSD), supernovae type Ia (SNIa) and Cepheids. Data analysis with CMB alone finds that the  today's Hubble parameter $H_0$ is consistent with its determination from Cepheids at $1\sigma$, resolving the famous tension of the cosmological standard models. The joint analysis of CMB, BAO, RSD and SNIa sets a lower bound for the sum of neutrino masses which is $\Sigma m_\nu >0.11$ eV at 1$\sigma$, in addition to the usual upper limit. 
The model selection analysis based on the effective $\chi_\text{eff}^2$ and Deviance Information Criterion is not able to clearly identify the statistically favored model between $\Lambda$CDM and the generalized cubic covariant Galileon, from which we conclude that the latter model deserves further studies.

\end{abstract}
\date{\today}

\maketitle

\section{Introduction}

The late time cosmic acceleration is one of the most puzzling phenomena in modern cosmology. Its modeling within  General Relativity (GR)  through  the cosmological constant ($\Lambda$) results in the $\Lambda$-cold-dark-matter (\lcdm) scenario. Although the latter gives a precise description of the Universe, it is known that it still contains a number of unresolved problems \cite{Joyce:2014kja}. These lead researchers to look for alternatives in the forms of an additional dark fluid, namely dark energy (DE) or modifying the gravitational law at cosmological scales, for example by including additional degrees of freedom (dofs), defining the so called modified gravity theories (MG) \cite{Lue:2004rj,Copeland:2006wr,Silvestri:2009hh,Capozziello:2011et,Clifton:2011jh,Tsujikawa:2010zza,Joyce:2014kja,Koyama:2015vza,Ferreira:2019xrr,Kobayashi:2019hrl}.  One of the most studied proposals of MG is the Horndeski theory (or Galileon theory)~\cite{Horndeski:1974wa,Deffayet:2009mn}, characterized by the presence of four free functions, namely $[G_2,G_3,G_4,G_5][\phi,X]$, where $\phi$ is the extra scalar field, whose dynamics is settled by second order equations of motion, and $X=\partial_\mu \phi\partial^\mu\phi$.  Recently, a proposal in cubic-order Horndeski theories, the Galileon ghost condensate model~\cite{Kase:2018iwp}, showed to be  statistically preferred over the standard \lcdm scenario due to a suppression in the integrated-Sachs-Wolfe (ISW) tail and a different behavior in the expansion history~\cite{Peirone:2019aua}. Another promising proposal is the generalized covariant Galileon model~\cite{DeFelice:2011bh}, which extends the covariant Galileon~\cite{Deffayet:2009wt} by considering in the Lagrangians power laws functions of $X$ ($G_i \propto X^{p_i}$, where $p_i$ are free constant parameters).  The chosen form of the $G_i$ functions allows for the existence of tracker solutions~\cite{DeFelice:2010pv,DeFelice:2010nf,DeFelice:2011bh,Frusciante:2018tvu}. This model has a viable parameter space, free from ghosts and Laplacian instabilities~\cite{DeFelice:2011bh}. 
Cosmological constraints at the background level show that the DE equation of state $w_{\rm DE}$ can take values very close to $-1$, allowing for the tracker to mimic \lcdm~\cite{DeFelice:2011aa}. Furthermore, the additional freedom given by the parameters $p_i$ might overcome the large enhancement of perturbations of the  covariant Galileon model which is proven to be disfavored by cosmological measurements~\cite{Renk:2017rzu,Peirone:2017vcq,Leloup:2019fas}. 

After the multi-messenger observation of the binary neutron stars merger event GW170817~\cite{TheLIGOScientific:2017qsa,Coulter:2017wya,GBM:2017lvd}, all MG models which predict a modification in the speed of propagation of gravitational waves (GWs) larger than  $10^{-15}$ are strongly disfavored~\cite{Creminelli:2017sry,Ezquiaga:2017ekz,Baker:2017hug,Sakstein:2017xjx,Bettoni:2016mij,Kase:2018aps}. In detail the Quintic Horndeski Lagrangian is ruled out and $G_4(\phi)$ reduces to be a standard conformal coupling to the Ricci scalar~\cite{Creminelli:2017sry,Baker:2017hug}. Applying the GWs constraint  to the generalized covariant Galileon model, it further restricts the Lagrangians to contain solely the K-essence Lagrangian ($G_2$), the  Cubic one ($G_3\Box\phi$) and a standard Einstein-Hilbert term. Hereafter, we will refer to this model as Generalized cubic covariant Galileon (GCCG). The GCCG model keeps the tensor speed unchanged and, additionally, a previous cosmological analysis shows a positive ISW-Galaxy cross-correlation~\cite{Giacomello:2018jfi}, contrary to what found for other cubic covariant Galileon models~\cite{Barreira:2014jha,Renk:2017rzu,Kobayashi:2009wr}. 

The aim of the present work is to extend previous studies on GCCG and perform a thorough investigation of its phenomenology at linear level. To this purpose we will make use of the effective field theory (EFT) of dark energy formalism~\cite{Gubitosi:2012hu,Bloomfield:2012ff} and its implementation in the Einstein-Boltzmann code  \eftcamb \cite{Hu:2013twa,Raveri:2014cka}. We will also provide cosmological constraints on the model and cosmological parameters at both background and linear level using present day data. An additional novelty in the analysis will be the inclusion of massive neutrinos. The latter can be constrained using cosmological data as they leave precise and measurable effects on cosmological observables~\cite{Lesgourgues:2006nd,Wong:2011ip}. Since such effects are similar to those observed in DE and MG scenarios: thus we will investigate the degeneracy between massive neutrinos and the GCCG model. 
Furthermore, we will use appropriate combinations of datasets in order to explore whether the GCCG model can ease the tension arising within the $\Lambda$CDM model between cosmic microwave background radiation (CMB) measurements and  the local estimate of the present day Hubble constant ($H_0$). The significance of such tension is rather high ($4.4 \sigma$) when comparing Planck with measurements of $H_0$ based on the cosmic distance ladder~\cite{Riess:2011yx,Riess:2016jrr,Riess:2019cxk}. Baryon acoustic oscillations (BAO) measurements from BOSS and SDSS show a $2.5 \sigma$ discrepancy in $H_0$ with Planck~\cite{Delubac:2014aqe}. Such tension is reduced by DES measurements~\cite{Abbott:2017wau,Abbott:2017smn}  and by calibration of the tip of the red giant branch applied to SNIa in the Large Magellanic Cloud \cite{Freedman:2019jwv}. We notice that in the latter case a different estimation of the Large Magellanic Cloud extinction can yield again to a large discrepancy \cite{Yuan:2019npk}. Phenomenological  DE and MG models seem to be very promising in reducing this tension \cite{Karwal:2016vyq,Pourtsidou:2016ico,Poulin:2018cxd,Lin:2018nxe,Agrawal:2019dlm,Agrawal:2019lmo,Kaloper:2019lpl,Desmond:2019ygn,DiValentino:2019jae,Sakstein:2019fmf}. 
Alternatively, it has been argued that the discrepancy can be due to the impact of the local density inhomogeneity on the calibration of SNIa distances with the Cepheids and the anchors \cite{Lombriser:2019ahl}.

The manuscript is organized as follows. In Sec.~\ref{Sec:model} we present the GCCG model and the procedure we adopt to solve the corresponding  background equations. In Sec.~\ref{Sec:Method}, we illustrate the methodology: we introduce the EFT formalism and we derive the mapping relations needed for the implementation in \eftcamb. We also present the stability analysis of the model and its departures from \lcdm in the cosmological observables. Then, in Sec.~\ref{Sec:Constraints} we present the cosmological datasets used for the Markov Chain Monte Carlo (MCMC) analysis and we show the results. Finally, we conclude in Sec.~\ref{Sec:conclusion}.

%%%%%%%%%%%%%%%%%%%%%%%
\section{The model}\label{Sec:model}
%%%%%%%%%%%%%%%%%%%%%%%

Let us consider the  cubic Horndeski action~\cite{Deffayet:2010qz,Kobayashi:2010cm}
\begin{eqnarray}\label{action1}
S=\int d^4x\sqrt{-g}\l(\f{m_0^2}{2} R +L_2+L_3\r)+ S_m[g_{\mu\nu},\chi_i],
\end{eqnarray}
with
\begin{eqnarray}
{ L}_2=G_2(\phi, X), \quad { L}_3=  G_3(\phi, X)  \Box\phi,
\end{eqnarray}
where $m_0^2$ is the Planck mass and $R$ is the Ricci scalar, $g_{\mu\nu}$ is the metric and $g$ is its determinant. $S_m$ is the matter action for all matter fields, $\chi_i$. 

On a flat Friedmann-Lema\^{i}tre-Robertson-Walker (FLRW) background of the form
\be
ds^2=a(\tau)^2\l(-d\tau^2+\delta_{ij}dx^idx^j\r) \,,
\ee
where $a(\tau)$ is the scale factor and $\tau$ is the conformal time, the Friedmann equations associated to the action \eqref{action1} are:
\ba
3m_0^2\hub^2&=&a^2(\rho_m+\rho_\phi)\,,\label{Fri1} \\
m_0^2(2\dot{\hub}+\hub^2)&=& -a^2(p_m+p_\phi)\,,\label{Fri2}
\ea
where $\hub=d\ln a/d\tau$ is the Hubble rate in conformal time, dot stands for derivatives with respect to $\tau$, $\rho_m$ and $p_m$ are the density and pressure of the matter fluids and 
\ba
&&\rho_\phi= 2XG_{2X}-G_2-6X\hub\phi^\prime G_{3X}-XG_{3\phi}\,,\\
&&p_\phi= G_2+2X\l(\f{\dot{\hub}}{a}\phi^\prime-\hub^2\phi^{\prime\prime}\r)G_{3X}-XG_{3\phi}\,,
\ea
are the density and pressure of the scalar field.  Here the prime is the derivative with respect to the scale factor, $G_{iX}=\partial G_i/\partial X$ and $G_{i\phi}=\partial G_i/\partial \phi$.  As usual we consider the continuity equations for the matter fields which we assume to be perfect fluids.
The equation of evolution for the scalar field is obtained by varying the action \eqref{action1} with respect to $\phi$ and at background level it reads
\be\label{phieq}
\f{\hub}{a^3}\f{d}{da}\l(a^3 J\r)=P\,,
\ee
and 
\ba\label{eq:phi}
&&J=-2 \hub \phi'G_{2X}-6\frac{ \hub }{a}X G_{3X}+2\hub \phi'G_{3\phi}\,,\\
&&P=G_{2\phi}-2X\l(G_{3\phi\phi}-2\l(\f{\dot{\hub}}{a}\phi^\prime-\hub^2\phi^{\prime\prime}\r)G_{3\phi X}\r)\,.
\ea

For the present analysis we consider the GCCG model~\cite{DeFelice:2011bh} specified by the  following forms
 of $G_2$ and $G_3$~\cite{DeFelice:2011bh}  \footnote{The notation slightly differs from that in ref.~\cite{DeFelice:2011bh} because we adopt  a different definition for X.}:
\ba
G_2=-c_2  \alpha _2^{4 \left(1-p_2\right)} \l(-X\r)^{p_2},\quad G_3=-c_3  \alpha _3^{1-4 p_3}  \l(-X\r)^{p_3}\,,
\ea
with $c_i, \alpha_i,p_i$ being constants, in particular 
\be
\alpha_2=\sqrt{H_0 m_0}\,,\quad \alpha_3=\left(\frac{m_0^{1-2 p_3}}{H_0^{2 p_3}}\right){}^{\frac{1}{1-4 p_3}}\,, 
\ee
where $H_0$ is the Hubble parameter at present time.  This model generalizes  the cubic covariant Galileon model (G3)~\cite{Deffayet:2009wt}. The latter is obtained in the limit $p_2=p_3=1$. Hereafter we will fix $c_2=1/2$ without loss of generality~\cite{Barreira:2013xea,2014JCAP...08..059B,Renk:2017rzu}.

The GCCG model shows a tracker solution given by~\cite{DeFelice:2011bh}
\be\label{tracker}
\l(\f{\hub}{a}\r)^{2q+1}\psi^{2q}=\zeta H_0^{2q+1}\,, 
\ee
where $\zeta$ is a dimensionless constant. For convenience we have introduced a dimensionless quantity:
\ba
 q=(p_3-p_2) +\frac{1}{2} \,,
\ea  
and a dimensionless scalar field:
\be
\psi=\f{1}{m_0}\f{d \phi}{d\ln a}\,.
\ee
Thus, for a fixed  $q$ the tracker attracts solutions with different initial conditions to a common trajectory.  

We solve the background equations for the GCCG model along the tracker solution, thus  the Friedmann equation \eqref{Fri1}  can be rewritten as
\be\label{Fried1}
\l(\f{\hub}{aH_0}\r)^{2+s}=\Omega_\phi^0+\l[\f{\Omega_c^0+\Omega_b^0}{a^3}+\f{\Omega_r^0}{a^4}+\Omega_\nu^0\f{\rho_\nu}{\rho_\nu^0}\r]\l(\f{\hub}{aH_0}\r)^{s}\,,
\ee
where $s=p_2/q$~\footnote{Let us note that the definition of the $s$ parameter in this work and that in ref.~\cite{Giacomello:2018jfi}, differs by a factor 2, i.e. $\tilde{s}=s/2$.}, $\Omega^0_{i}=\f{\rho_i^0}{3m_0^2H_0^2}$ are the density parameters at present time of the cold dark matter (c), baryons (b), radiation (r) and massive neutrinos ($\nu$) with density $\rho_\nu$. In the above equation we have also used the solutions of the continuity equations for the cold dark matter ($\rho_c=\rho_c^0/a^3$), baryons ($\rho_b=\rho_b^0/a^3$) and radiation ($\rho_r=\rho_r^0/a^4$).  We have also identified  the density of the scalar  field at present time as
\ba\label{flatness}
\Omega_\phi^0&=&1-\Omega_m^0=c_3 (2 sq+2 q-1) \zeta ^{s+1}-\frac{1}{6} (2 sq -1)\zeta ^{s}\,,
\ea
by considering  eq. \eqref{Fri1} at present time.

Once the Friedmann equation \eqref{Fried1} has been solved for $\hub$ the scalar field is completely determined trough the tracker solution eq.~(\ref{tracker}).

Now, it remains to consider the  equation for the scalar field. Using the tracker solution in eq. \eqref{phieq} and evaluating the latter at present time, we get a constraint equation
\be \label{const2}
-sq +3 c_3 \zeta  (2 sq +2 q-1)=0.
\ee 

We can combine the eq.~(\ref{flatness}) and the above constraint to eliminate two parameters, $\zeta$ and $c_3$, then:
\ba
\zeta=\l(6\Omega_\phi^0\r)^{\f{1}{s}}\,,\qquad c_3=\f{1}{3}\f{sq }{\l(6\Omega_\phi^0\r)^{\f{1}{s}}(2sq+2q-1)}\,.
\ea
We conclude, that the resulting GCCG model has two additional free parameters, i.e. $\{s, q\}$, with respect to $\Lambda$CDM.

%%%%%%%%%%%%%%%%%%%%%%%
\section{Methodology}\label{Sec:Method}
%%%%%%%%%%%%%%%%%%%%%%%%

%%%%%%%%%%%%%%%%%%%%%%%%%%%%%%%%%
\subsection{Effective field theory for dark energy}\label{Sec:EFT}
%%%%%%%%%%%%%%%%%%%%%%%%%%%%%%%%%

The primary goal of the present investigation is to  study the linear cosmological perturbations  and perform cosmological constraints of the GCCG model. In particular we will be interested in the evolution of scalar modes since the tensor modes are left unmodified with respect to GR. To this purpose  we will use the EFT approach~\cite{Gubitosi:2012hu,Bloomfield:2012ff,Gleyzes:2013ooa,Piazza:2013pua,Tsujikawa:2014mba,Frusciante:2019xia}. The advantage in using this approach relies on the possibility to use the publicly available \texttt{EFTCAMB}/\texttt{EFTCosmoMC} package~\cite{Hu:2013twa,Raveri:2014cka} \footnote{Web page: \url{http://www.eftcamb.org}} which with minor modifications allows to perform the desired analysis. In the following we will briefly discuss the necessary steps (see ref.~\cite{Hu:2014oga} for additional details).

Within the EFT framework it is possible to write the linear perturbed action around a flat FLRW background and in unitary gauge for all DE and MG with one additional scalar dof. Here, we will consider the restriction of the EFT action to the subclass of Horndeski theory with luminal propagation of tensor modes which  reads 
\begin{align}\label{eftaction}
\mathcal{S} =& \int d^4x \sqrt{-g} \frac{m_0^2}{2} \bigg\{  \left[1+\Omega(\tau)\right] R + \f{2\Lambda(\tau)}{m_0^2} - \f{2c(\tau)}{m_0^2}a^2\delta g^{00}  \nonumber\\
  & + H_0^2 \gamma_1 (\tau) \left( a^2\delta g^{00} \right)^2
 - H_0 \gamma_2(\tau) \, a^2\delta g^{00}\,\delta K\bigg\}, 
\end{align}
where  $\delta g^{00}$ and $\delta K$  are  the perturbations respectively of the upper time-time component of the metric and the trace  of the extrinsic curvature. $\{\Omega, c, \Lambda,\gamma_1,\gamma_2\}$ are the so called EFT functions.  $\Lambda$ and $c$ can be expressed in terms of $\Omega$, $\hub$ and the densities and pressures of matter fluids by using the background field equations~\cite{Gubitosi:2012hu,Bloomfield:2012ff}, thus we are left with three free functions of time.  The latter can be specified  following the mapping procedure, i.e. given a specific covariant theory it can be rewritten in the EFT language as discussed in details in refs.~\cite{Gubitosi:2012hu,Bloomfield:2012ff,Bloomfield:2013efa,Gleyzes:2013ooa,Gleyzes:2014rba,Frusciante:2015maa,Frusciante:2016xoj}.  For the GCCG model using  the tracker solution (\ref{tracker}) we obtain:
\ba
&&\gamma_1=  \f{s}{4}  \Omega^0_{\phi }  \l(\f{aH_0}{\hub}\r)^{s}  \left[12 q^2+1 -\f{\dot{\hub}}{\hub^2}\right]\,,\\
&&\gamma_2=-2sq \Omega^0_{\phi} \l(\f{aH_0}{\hub}\r)^{1+s}  \,,
\ea
and  $\Omega=0$. 
The above EFT functions will be fully specified once the eq. \eqref{Fried1} is solved. We have implemented the above mapping and the background solver described in Sec.~\ref{Sec:model} into \eftcamb. After these modifications, the code evolves the linear perturbation equations for the GCCG model and computes the relevant linear cosmological observables.  

At the level of perturbations only the $\gamma_2$ functions will alter the cosmological observables. This modification is related to the braiding effect which is due to the mixing between the metric and the DE field \cite{Deffayet:2010qz}. From the above relation we notice that both $s$ and $q$ have a relevant role. 
On the contrary $\gamma_1$ has no measurable effect  being its contribution to the observables below the cosmic variance \cite{Frusciante:2018jzw}.
However it is important in defining the viable parameter space. 

%%%%%%%%%%%%%%%%%%%%%
\subsection{Stability}\label{Sec:stability}
%%%%%%%%%%%%%%%%%%%%%

A physically viable theory needs to satisfy specific requirements:  the no-ghost condition to  prevent the development of dofs with a negative kinetic term, the no-gradient condition to avoid the presence of modes with a negative speed of propagation, $c_s$, and  the absence of tachyonic instabilities which appear when  the perturbations are not computed about the true vacuum of the theory. Both the no-ghost and no-gradient conditions are high momenta ($k$) statements while the tachyonic instability is relevant at low-$k$. So that, they identify a theoretically rigorous set of conditions that guarantees stability of the theory at all cosmological scales.  When studying cosmological perturbations, the matter fields and their mixing with the scalar field cannot be neglected as they can change the viability space of the theory. For example in the Gleyzes-Langlois-Piazza-Vernizzi theory \cite{Gleyzes:2014qga} the scalar and the matter fields do not decouple at high-$k$ and their speeds of propagation are modified \cite{Gleyzes:2014qga,Gergely:2014rna,Kase:2014yya,DeFelice:2015isa,DeFelice:2016ucp}.  In Horndeski theory the matter fields are involved only in the tachyonic condition \cite{DeFelice:2016ucp} and its impact on the viability space has been widely investigated \cite{Frusciante:2018vht}.  Since this theory introduces an extra scalar dof, the new no-ghost condition associated to it, leads also, automatically, to an additional condition, that we will discuss later on, meant to avoid strong coupling problems.

In this Section we will discuss the theoretical viability requirements which guarantee the GCCG model does not develop any pathological instability. They are: 
\begin{itemize}
\item \textit{no-ghost condition:}   In order to find such condition, we choose to study the kinetic term for the field $\phi=\phi(\tau)+\delta\phi$, and we define $Q_s$ so as to have $\mathcal{L}\ni 1/2\,a^2\,Q_s\,\dot{\delta\phi}^2$ \footnote{More in detail, after choosing a gauge (e.g.\ the flat gauge, but this choice in not special), adding two fluids to model matter fields (representing the relativistic and dust matter fields), and removing all the auxiliary fields, we can diagonalize the three by three kinetic matrix, in order to find the no-ghost conditions. This can be done my making field redefinitions for the matter fields $\delta_{m,r}$ (such as $\delta_m=\delta_{m}^{\rm new}+C(\tau)\,\delta\phi$, where $C$ is chosen as to diagonalize the kinetic matrix). Then, two out of three no-ghost conditions refer to the matter fields and are trivially satisfied. On the other hand the no-ghost condition for the scalar field $\delta\phi$, in the high-$k$ regime, requires $Q_s>0$.} Its expression reads \cite{Giacomello:2018jfi}
  \begin{equation}
    Q_s={\frac {6 \left( \Omega_{{\phi}}s+2 \right) \Omega_{{\phi}}s{q}^{2}(\hub/a)^{2}m_0^{2}}{(\dot\phi/a)^{2} \left( \Omega_{{\phi}}sq-1 \right) ^{2}}}>0\,,
\end{equation}
where $\Omega_{\phi}\equiv a^2\rho_{\phi}/(3m_0^2\hub^2)$, which implies $s>0$.
\item \textit{no-strong coupling condition:} 
 In order to avoid $Q_s \rightarrow 0$ at early times, we also require  \cite{Giacomello:2018jfi}
\be
\f{qs-1}{q(2+s)}\le 0\,,
\ee
which follows from $Q_s\propto\Omega_{\phi}^{(qs-1)/[q(2+s)]}$.

\item \textit{no-gradient condition:}  the avoidance of any potential gradient instabilities at high-$k$ requires a positive  speed of propagation, $c_s^2>0$. For the model under consideration it gives \cite{DeFelice:2016ucp}
\ba
c_s^2&=& 
\f{(2q(s+2)-1)\l(1-\f{\dot{\hub}}{\hub^2}\r)-2q^2s\Omega_\phi^0\l(\f{aH_0}{\hub}\r)^{s+2}}{6q^2\l(1+\f{s}{2}\Omega_\phi^0\l(\f{aH_0}{\hub}\r)^{s+2}\r)}>0\,.\,\,\,\,
\ea

\item \textit{no-tachyonic condition or mass condition:}  the tachyonic instability occurs when the  Hamiltonian is unbounded from below at low-$k$. This occurs  when the negative mass eigenvalues of the Hamiltonian are much larger, in absolute value, than the Hubble parameter \cite{DeFelice:2016ucp}, i.e. 
\be\label{mass}
|\mu_i(\gamma_1,\gamma_2, \rho_m)|< \beta^2\f{\hub^2}{a^2}\,,
\ee
where the eigenvalues $\mu_i$ are expressed in terms of the EFT functions and matter fields and $\beta$ is a constant defining the rate of allowed instability. Because of the length and complexity of the expressions of the mass eigenvalues we refer the reader to ref.~\cite{DeFelice:2016ucp}, where these conditions have been derived. Among the stability requirements the mass condition is in general  the less severe \cite{Frusciante:2018vht}. Eq.~\eqref{mass} in some cases can be quite conservative for $\beta=1$ since  there might be cases in which the instability does not occur even requiring $|\mu_i|<10^2\hub^2/a^2$. In the following we will investigate the impact of such condition on the viable parameter space.
 \end{itemize}
%-----------------------------------------------------------------------
\begin{figure}[t!]
\centering
\includegraphics[width=0.3\textwidth]{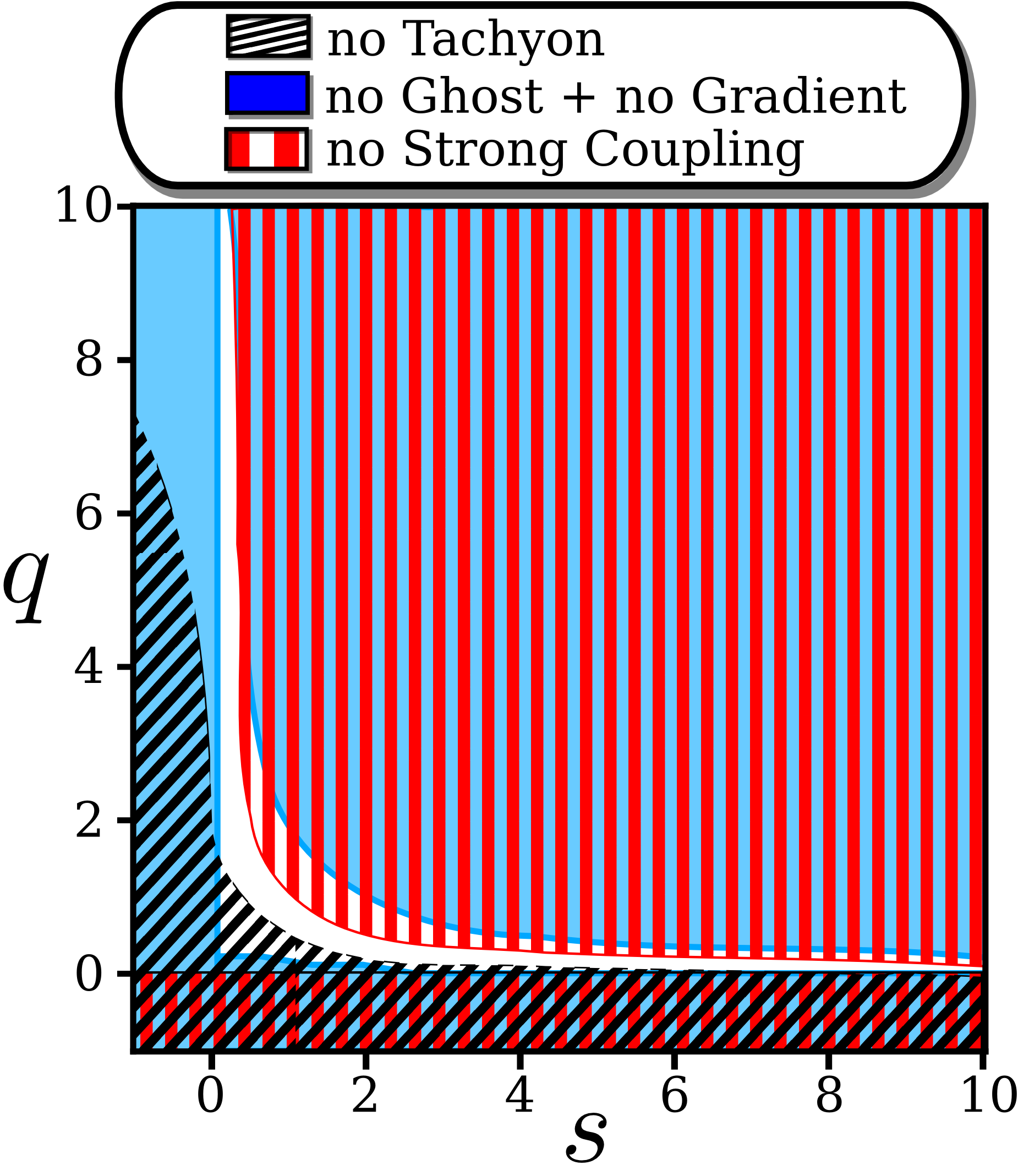}
\caption{\label{fig:stability}  The stable parameter space of GCCG is shown in white, while different filled regions represents cuts due to the stability conditions, discussed in Sec.~\ref{Sec:stability}, as shown in the legend. For the mass condition we have set $\beta=1$.}
 \end{figure}
%-----------------------------------------------------------------------
 \begin{figure*}[t!]
\centering
\includegraphics[width=0.9\textwidth]{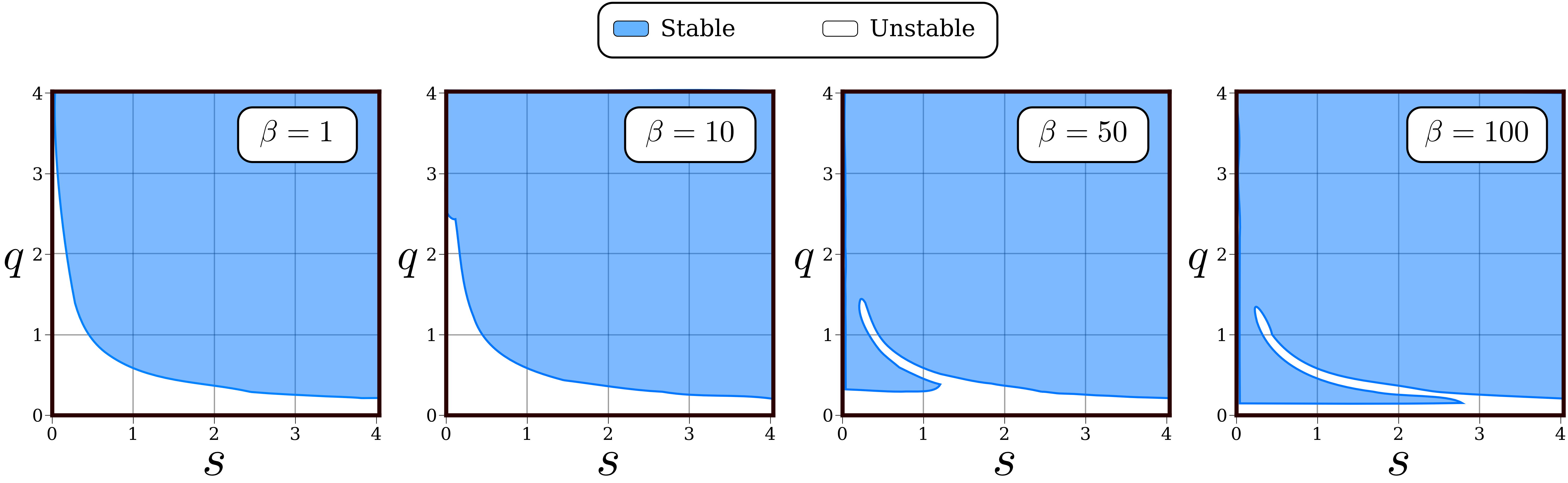}
\caption{\label{fig:stabcuts}  We show the impact of the mass condition on the parameter space of the GCCG model for different values of $\beta$. The stable parameter space is shown in blue, while the region undergoing mass instability is white.   For these plots we choose four different mass instability rates $\beta=1,10, 50,100$.}
 \end{figure*}
%-----------------------------------------------------------------------

Let us note that the above conditions apply only to the scalar sector because the tensor modes are not modified neither in the speed of propagation (by construction $c_t^2=1$ \cite{Deffayet:2010qz}) nor in the kinetic coefficient $(Q_t=1/2)$. These conditions are implemented in \eftcamb and  are used as \textit{viabiliy priors} in \texttt{EFTCosmoMC}. 
 
In Figure~\ref{fig:stability} we show the effects of the different stability filters on the $s$-$q$ parameter space.
The no-ghost, no-gradient  and strong coupling conditions identify the same parameter space as in ref.~\cite{Giacomello:2018jfi} and we also recover the hyperbole $sq=1$ given by the strong coupling condition which separates the stable and unstable region. Here we  also include the cut of  the mass condition for $\beta=1$ which modifies the parameters space for small values of both $q$ and $s$. In Figure~\ref{fig:stabcuts} we show the different shapes of the stable parameter space when different mass cuts are applied. These correspond to different choices of the $\beta$ parameter.  We notice that the parameter space does not change for any value of $\beta\leq10$. Larger values of $\beta$ allow for an extended viable region with a peculiar shape. In particular we see that for a very  large value of $\beta$ ($\beta=100$) a small unstable area within the stable one is identified.  We have verified that  values of $q$ and $s$ within such  area would indeed lead to the evolution of unstable modes.  While large part of the parameter space cut off by $\beta=1, 10$ is actually stable.  

For this reason, when constraining the model parameters against data we will consider the case in which the mass condition is switched off because we prefer to sample the larger viable parameter space at our disposal, thus we consider as baseline stability conditions the no-ghost, no-gradient and strong coupling condition.  However, we will also discuss how the constraints on the cosmological and model parameters will change if the mass condition (with $\beta=1$) is included on top of the baseline conditions.

%%%%%%%%%%%%%%%%%%%%%%%%%%
\subsection{Cosmological implications}\label{Sec:cosmology}
%%%%%%%%%%%%%%%%%%%%%%%%

%-----------------------------------------------------------------------
\begin{center}
\begin{table}
\begin{tabular}{|l|c|c|c|}
\hline
Model &\,\,\, $s$\,\,\,& $q$ & $\sum m_\nu$ (eV)  \\
\hline
\hline
G3          & 2 & 0.5  & --    \\
G3$+\nu$ &2  & 0.5  & 0.85   \\
GCCG1& 2& 0.35 & --  \\
GCCG1$+\nu$ & 2 & 0.35 & 0.85 \\
GCCG2 & 1.3 & 0.5 & --\\
GCCG2$+\nu$ &1.3 & 0.5 & 0.85  \\
\hline
\end{tabular}
\caption{Values of $s$ and $q$ for the Cubic Galileon (G3) and the two GCCG models presented here. For each case we also consider a cosmology with massive neutrinos. The sum of the neutrino masses adopted is the $1\sigma$ constraint for G3 obtained in ref.~\cite{Peirone:2017vcq}. The standard cosmological parameters are chosen to be: $\Omega_b^0\,h^2=0.0226$, $\Omega_c^0\,h^2=0.112$  with $h=H_0/100$ and $H_0=70 \,\mbox{km}/\mbox{s}/\mbox{Mpc}$. All the values satisfy the theoretical conditions discussed in Sec.~\ref{Sec:stability}.}
\label{table:parameters}
\end{table}
\end{center}
%-----------------------------------------------------------------------

We will now perform a thorough analysis of the cosmological implications in the GCCG model. Let us introduce the perturbative flat FLRW metric written in Newtonian gauge 
\be
ds^2=a(\tau)^2[-(1+2\Psi)d\tau^2+(1-2\Phi)dx^2]\,,
\ee
where $\{\Psi(\tau,x^i), \Phi(\tau,x^i)\}$ are the gravitational potentials. For MG models the Poisson and lensing equations can be written in Fourier space as follows~\cite{Bean:2010zq,Silvestri:2013ne}:
\ba \label{mudef}
&&-k^2\Psi=4\pi G_N a^2\mu(a,k)\rho_m\Delta_m\,, \\
&&\label{sigmadef}
-k^2(\Psi+\Phi)=8\pi G_Na^2\Sigma(a,k)\rho_m\Delta_m\,,
\ea
where $G_N$ is the Newtonian gravitational constant, $\Delta_m$ is the total matter density contrast, $\mu$ and $\Sigma$ define respectively the effective gravitational coupling and the light deflection. The GR limit is recovered when $\{\mu,\Sigma\}=1$. In general MG models are characterized by an anisotropic stress term given by $\Phi\neq\Psi$.  
Because the GCCG model does not have any modification in the speed of propagation of GWs nor a running Planck mass, there is no anisotropic stress term and the two gravitational potentials are equal. From this follows that $\mu \simeq\Sigma$. 

Using the quasi-static approximation (QSA) and for sub-horizon perturbations it is possible to explicitly write the functional forms of $\mu$  for the GCCG model which reads 
\ba\label{QS_expressions}
\mu(a,k)= 1+ \f{s^2q^2(\Omega_\phi^0)^2\l(\f{aH_0}{\hub}\r)^{2(s+2)}}{Q_sc_s^2\l(1-sq\Omega_\phi^0\l(\f{aH_0}{\hub}\r)^{s+2}\r)^2}\,.
\ea
According to the above relation,  $\mu \geq 1$ for any viable value of $q$ and $s$ and hence the gravitational interaction is  always stronger than in GR.  We expect  modifications in the lensing potential ($\Phi+\Psi$) and thus in the  ISW effect, being the latter sourced by $\dot{\Psi}+\dot{\Phi}$, and  finally in the growth of structures. While Eq. \eqref{QS_expressions} is very useful to  grab some preliminary information about the physics of the model, in the following we will not rely on the QSA but we will solve the complete set of linear perturbation equations.

We analyze the dynamics of linear cosmological observables and quantify the deviation with respect to the standard scenario. We will always show for reference the \lcdm  and G3 evolutions. In Table. \ref{table:parameters} we list the parameters defining the models. They are chosen such that GCCG1 shares the same background evolution of G3 but a different value for $q$, while GCCG2 evolves differently at the background level while having the same value for $q$ as in G3.  We also include the cases with and without massive neutrinos. The sum of neutrino mass is chosen to be $0.85\,\, eV$  which is the $1\sigma$ constraint for G3 obtained in~\cite{Peirone:2017vcq}.

%-----------------------------------------------------------------------
\begin{figure}[t!]
\includegraphics[width=0.4\textwidth]{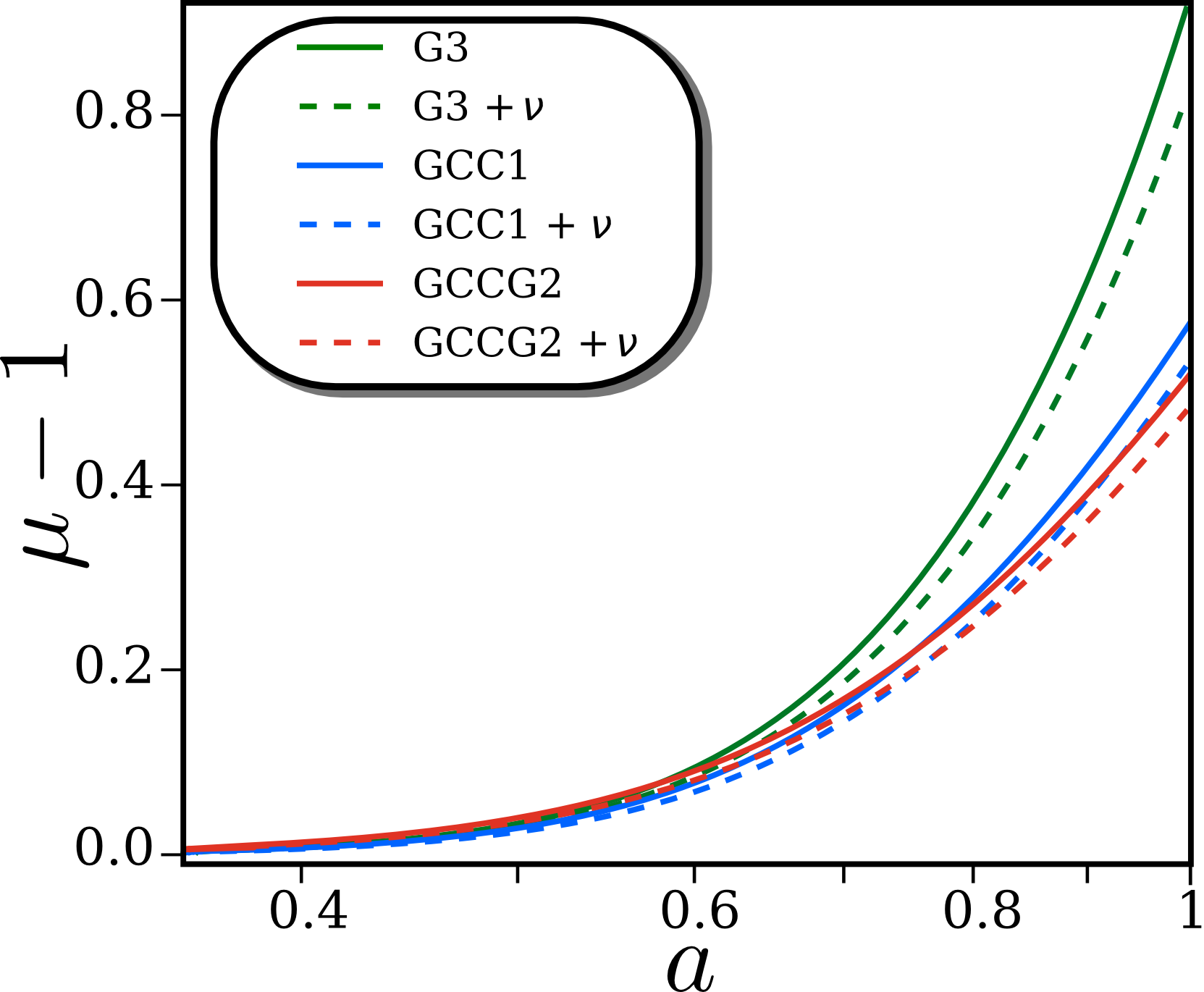}
	\caption{\label{fig:mu} Evolution of the effective gravitational coupling $\mu$ as function of the scale factor $a$ at $k=0.01$ Mpc$^{-1}$ for the test models in Table \ref{table:parameters}.} 
\end{figure}
%-----------------------------------------------------------------------

In Figure~\ref{fig:mu} we show the evolution of the effective gravitational coupling at $k=0.01$ Mpc$^{-1}$. We notice deviations from GR only at late times for $a>0.4$. As already mentioned, for all the cases considered we recover $\mu>1$. The largest deviation is obtained for G3 which reaches today $\mu^0-1=0.91$. It is followed by GCCG1 with $\mu^0-1=0.57$ and finally GCGG2 with $\mu^0-1=0.52$. We note that in the range $0.4<a<0.7$ the effective gravitational coupling for GCGG2 is larger than the one for GCCG1. It is only for $a>0.7$ that  $\mu(GCCG1)$ rapidly grows more.  The inclusion of the massive neutrinos lowers the deviation  with respect to \lcdm of about $4\%$.

%-----------------------------------------------------------------------
\begin{figure*}[t!]
\includegraphics[width=0.99\textwidth]{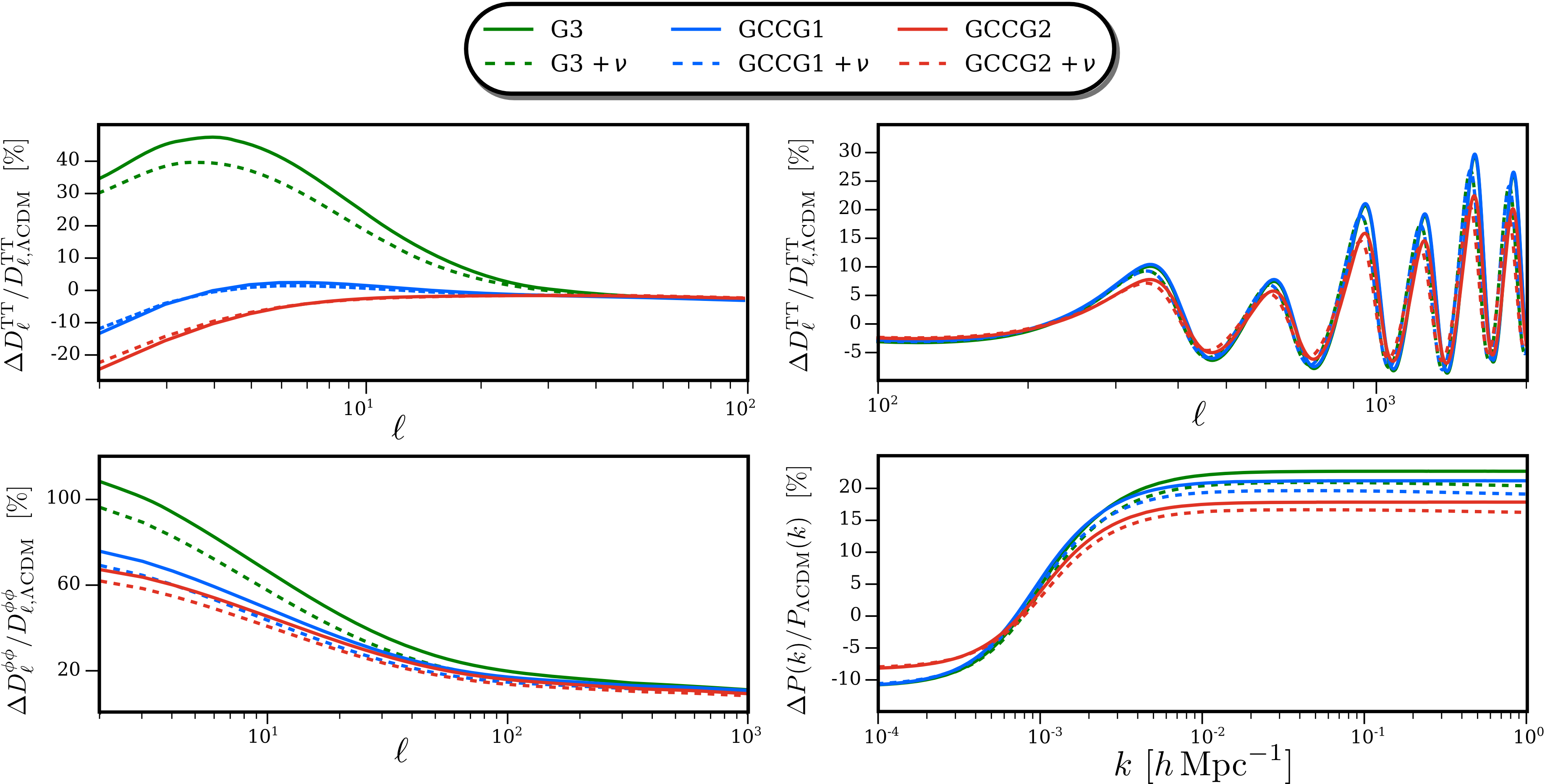}
	\caption{\label{fig:observables} Percentage differences of the test models in Table~\ref{table:parameters} relative to \lcdm in the power spectra. \textit{Upper panel}: Differences in the CMB temperature-temperature power spectrum  $D_\ell^{\rm TT} = \ell(\ell+1)C_\ell^{\rm TT}/(2 \pi)$  at low multipoles (left) and  large angular scales (right). \textit{Bottom panel}: Differences in the lensing potential auto-correlation power spectra $D_\ell^{\phi \phi} = \ell(\ell+1)C_\ell^{\phi \phi}/(2 \pi)$ (left) and matter power spectra $P(k)$ (right). } 
\end{figure*}
%-----------------------------------------------------------------------

In Figure~\ref{fig:observables} we show the differences relative to \lcdm for the CMB temperature-temperature (TT), lensing potential auto-correlation and matter power spectra. 
A different evolution of the gravitational potentials leads to modifications in the gravitational lensing power spectrum.  In the bottom left panel of Figure~\ref{fig:observables} we notice that the lensing power spectra for the modified cosmologies show an enhancement with respect to the $\Lambda$CDM scenario. This is expected as $\Sigma (\simeq\mu)>1$.  In both the GCCG models however the fluctuations of this observable are suppressed  for $\ell<200$ with respect to G3. The deviations  relative to $\Lambda$CDM are around 20\% for $\ell >100$ and grow up for  smaller values of $\ell$. In particular they reach $\sim 68\%$ for GCCG2, $\sim 80\%$ for GCCG1 and larger than 100\% for the G3 model.

\begin{figure}[t!]
\includegraphics[width=0.4\textwidth]{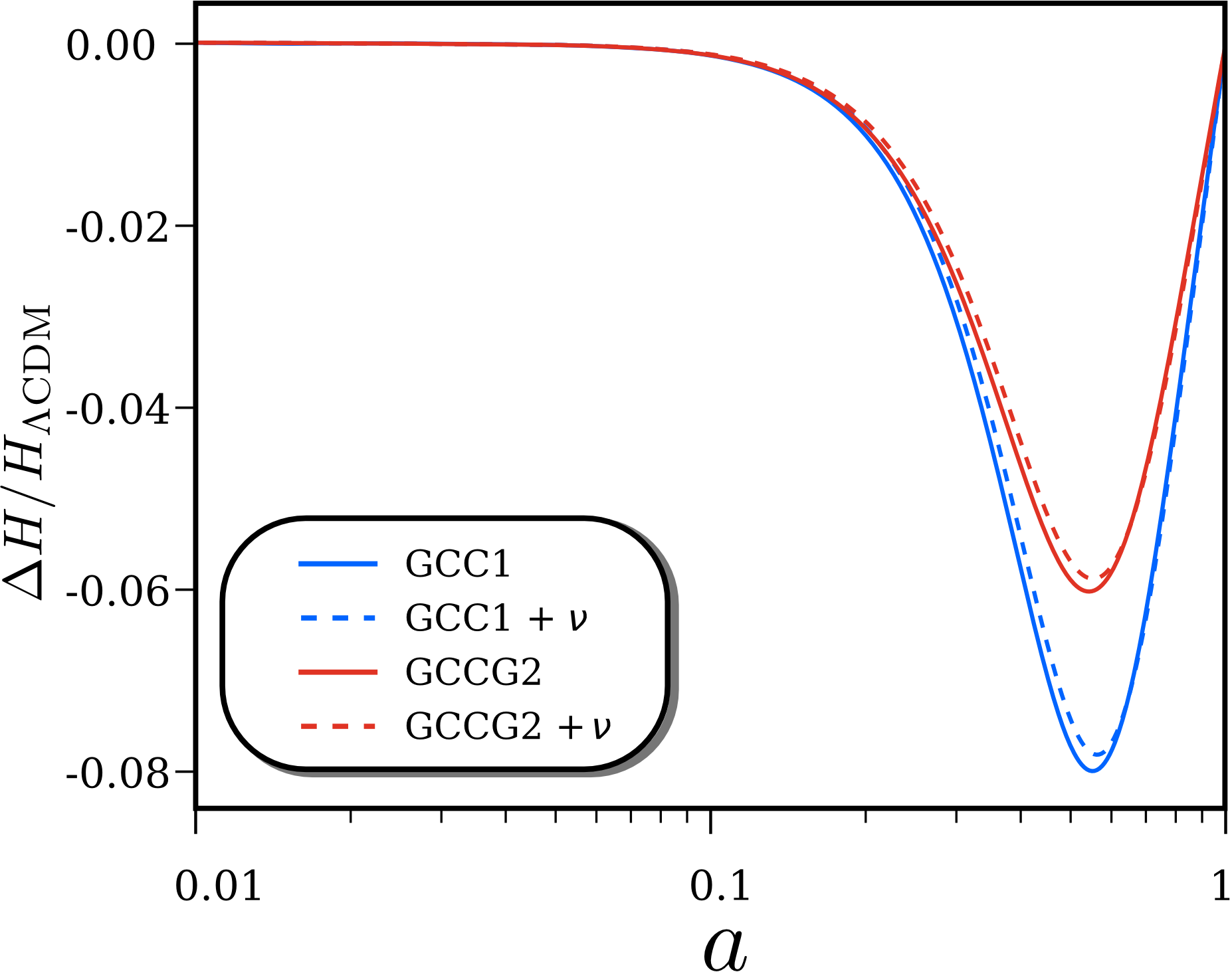}
	\caption{\label{fig:hubble} Relative difference in the expansion history of the test models in Table \ref{table:parameters} with respect to $\Lambda$CDM. G3 overlaps with GCCG1 as they share the same background evolution. } 
\end{figure}

 In the upper left panel we note that both the GCCG models  can predict an  ISW tail suppressed with respect  to G3 and for appropriate  values of the parameters even  with respect to the $\Lambda$CDM model.  
 GCCG1 shares the same background expansion history of G3, however a lower value of $q$ can suppress the ISW tail of about $15\%$ with respect to $\Lambda$CDM. Changing $s$ toward smaller values it is also possible to lower the low-$\ell$ tail in the TT power spectrum, for example in  GCCG2 the suppression with respect to \lcdm reaches the $28\%$. 
In the upper right panel we show the difference with respect to \lcdm in the TT power spectra for large angular scales. The modification of the gravity force  shifts the peaks and troughs to higher multipoles with respect the $\Lambda$CDM. This effect is mostly due to a change in the expansion history which alters the distance to the last scattering surface.  We note that the shift in the peaks for G3 and GCCG1 is $\sim 30\%$ while in the GCCG2 is $<25\%$.   The larger the shift  the wider is the deviation in the background evolution compared to \lcdm as shown in Figure~\ref{fig:hubble}. Finally, in the bottom right panel we show the relative difference in the matter power spectra.  We observe an enhancement of the growth of structure between  10\% and 20\% for $k>10^{-3}$ h\,Mpc$^{-1}$ and a suppression for very small $k$ which for the GCCG1 model is $\sim 10\%$ . 
 
Regardless of the observable we consider, the impact of the massive neutrinos goes in the direction of suppressing the MG effects as already noticed in Figure~\ref{fig:mu}.  Thus massive neutrinos push the GCCG model toward $\Lambda$CDM.

From this analysis we can deduce that the GCCG model shows a very interesting phenomenology.  The additional freedom given by $s$ and $q$ generates MG effects on observables that can be less strong than G3 ones. This feature might allow the model  to fit the data better than G3. The latter has been ruled out at $7.8\sigma$ using ISW data \cite{Renk:2017rzu}   afterwards such result was confirmed by a Bayesian model comparison involving several datasets \cite{Peirone:2017vcq}. Furthermore, the possibility to have a suppressed ISW tail with respect to $\Lambda$CDM might provide a better fit to data 
and drive model selection criteria toward the preference of GCCG over \lcdm, as already noticed for the Galileon ghost condensate model~\cite{Peirone:2019aua}.

%-----------------------------------------------------------------------
%-----------------------------------------------------------------------

%%%%%%%%%%%%%%%%%%%%%%%%%%%%
\section{Cosmological constraints and model selection}\label{Sec:Constraints}
%%%%%%%%%%%%%%%%%%%%%%%%%%%%

\subsection{Data sets} \label{Sec:Dataset}

In the present cosmological analysis, we employ the Planck 2015 measurements~\cite{Aghanim:2015xee,Ade:2015xua} of CMB temperature and polarization on large angular scales, limited to multipoles $\ell < 29$ (low-$\ell$ TEB likelihood) and the CMB temperature on smaller angular scales (PLIK TT likelihood, $30 < \ell < 2508$). We also vary the nuisance parameters used to model foreground as well as instrumental and beam uncertainties.  Given the similarities between the 2015 and 2018 releases, we do not expect that our results would change significantly if we employed the data presented in~\cite{Aghanim:2018eyx}. We complement the $Planck$ dataset with measurements of baryon acoustic oscillation (BAO) from the 6dF galaxy survey~\cite{Beutler2011}, the BAO scale measurements from the SDSS DR7 Main Galaxy Sample~\cite{Ross2015} and the combined BAO and redshift space distortion (RSD) data from the SDSS DR12 consensus release~\cite{Alam:2016hwk}. We also include data coming from the Joint Light-curve Array ``JLA" Supernovae (SNIa) sample, as introduced in~\cite{Betoule:2014frx}. We consider the above data sets in two combinations: $Planck$ alone and Planck+BAO+RSD+SNIa (hereafter $PBRS$). Where mentioned, we also consider a Gaussian prior on the Hubble constant $H_0 = 74.03\pm1.42$  km/s/Mpc, as estimated  in~\cite{Riess:2019cxk} using three anchors: cepheids in the Large Magellanic Cloud, the Milky Way cepheid parallaxes, and the masers in NGC 4258. We will  consider such measurement in combination with $Planck$  and we will refer to it as $Planck +H_0$.

For the MCMC likelihood analysis with the above data sets we use the \texttt{EFTCosmoMC} code \cite{Raveri:2014cka}. We impose flat priors on the two models parameters: $q \in [-10, 10]$ and $s \in [-10, 10]$ and we test that the results are insensitive on the choice of the prior volume.

%%%%%%%%%%%%%%%%%%%%%%
\subsection{Results and discussion}\label{Sec:Results}
%%%%%%%%%%%%%%%%%%%%%

%-------------------------------
\begin{figure}[t!]
\centering
\includegraphics[width=0.4\textwidth]{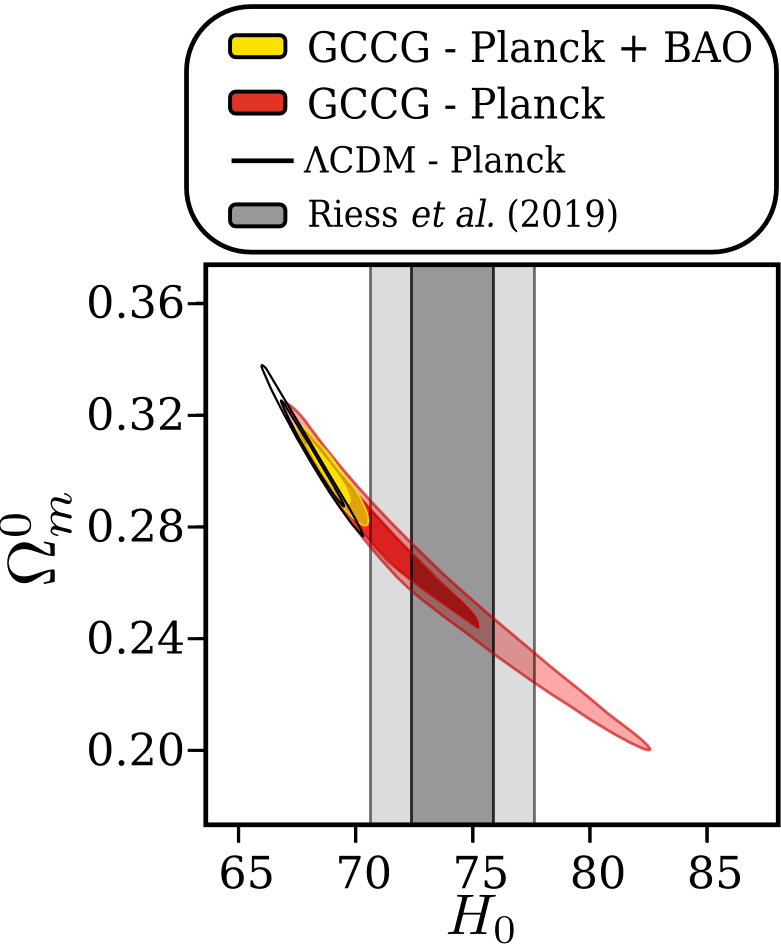}
\caption{\label{fig:tension}  The marginalized 2-D joint distribution for the cosmological parameters $\Omega_m^0$ and $H_0$, obtained from the analysis of the $Planck$ dataset with GCCG (in red) and with $\Lambda$CDM (in black). The inner region of the distribution represents the $68\%$ confidence level, while the outer region cuts the distribution at $95\%$. The vertical grey band represents the $1$ and $2$ $\sigma$ constraints obtained by the local measurements~\cite{Riess:2019cxk}. } 
 \end{figure}
 %-------------------------------
 \begin{figure}[t!]
\centering
\includegraphics[width=0.4\textwidth]{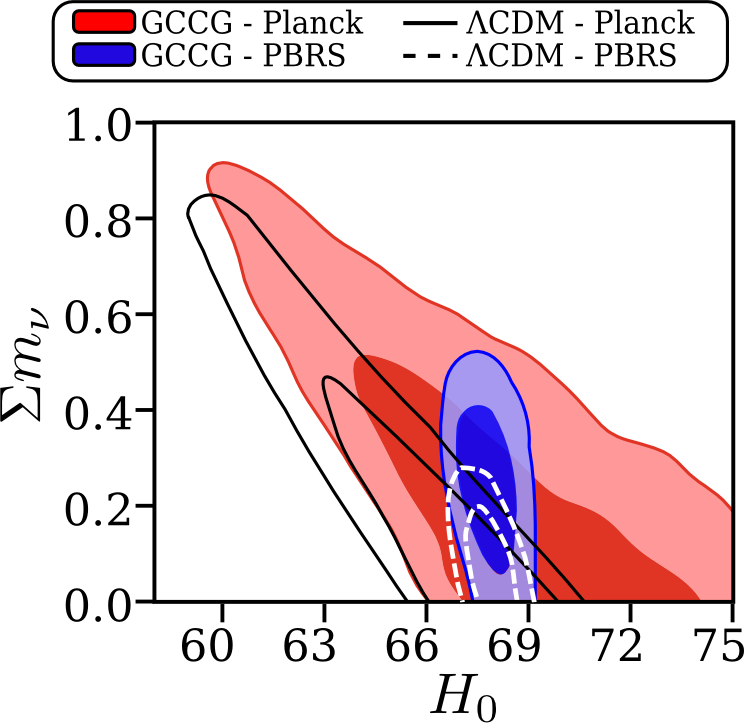}
\caption{\label{fig:H0mnu}  The marginalized 2-D joint distribution for  $H_0$ and $\sum m_\nu$, obtained from the analysis of  GCCG with the $Planck$ dataset (in red) and $PBRS$ (in blue).  The $\Lambda$CDM results are shown with solid black lines for $PBRS$ and dashed white lines for $Planck$. The inner region of the distribution represents the $68\%$ confidence level, while the outer region cuts the distribution at $95\%$.} 
 \end{figure} 
 %-------------------------------
 \begin{figure*}[t!]
\centering
\includegraphics[width=0.9\textwidth]{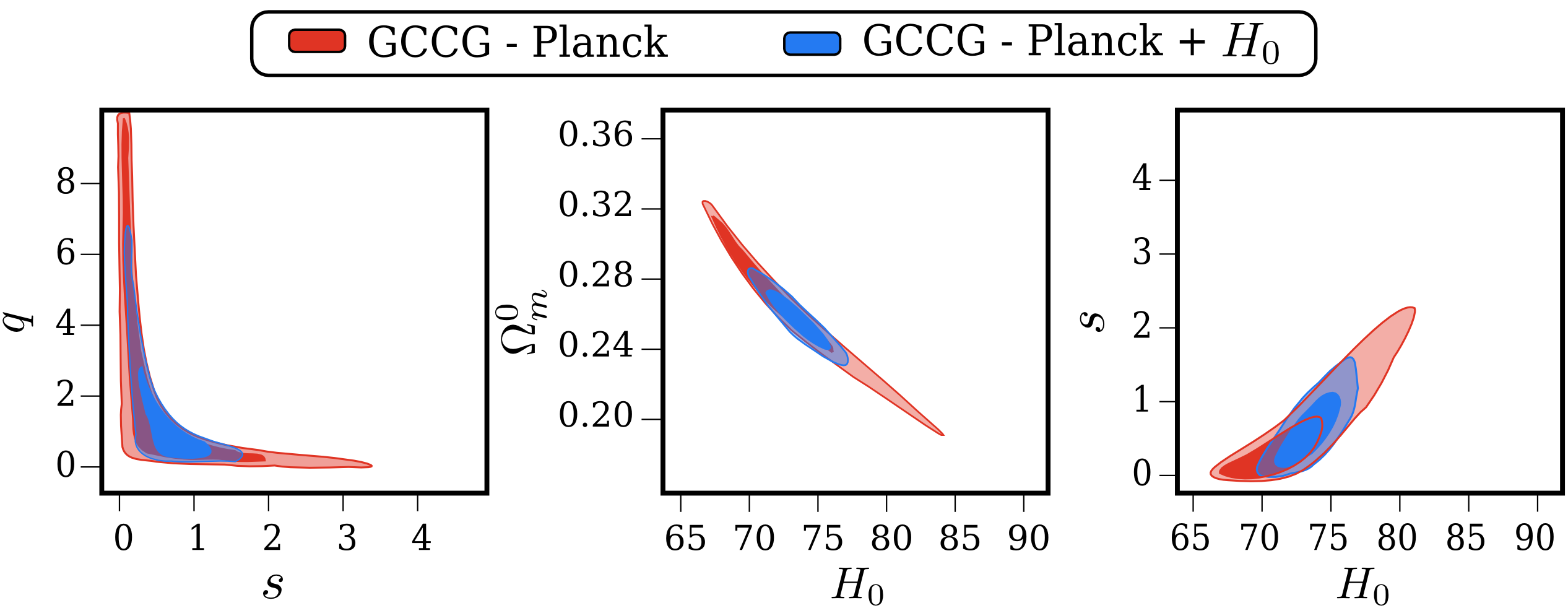}
\caption{\label{fig:degeneracysH0}  The marginalized 2-D joint distribution for the cosmological parameters $H_0$ and $\Omega_m^0$ and model parameters $S$ and $q$, obtained from the analysis of the $Planck$ dataset (in red). In blue we show the results of combining the CMB measurement with the local value of $H_0$~\cite{Riess:2019cxk}. The inner region of the distribution represents the $68\%$ confidence level, while the outer region cuts the distribution at $95\%$. } 
 \end{figure*}
%-------------------------------

In this Section we present and discuss the constraints of the cosmological and model parameters in the GCCG model for  two case studies: GCCG model with and without massive neutrinos. For the cosmological parameters we show the present day values of the matter density $\Omega_m^0$, Hubble parameter $H_0$,  the amplitude of the linear power spectrum at scale of 8 h$^{-1}$Mpc, denoted by $\sigma_{8}^0$ and the sum of neutrinos mass $\Sigma m_\nu$. We  include in Table \ref{tab_bestfit_params}, the marginalized constraints for the two combinations of datasets ($Planck$ and $PBRS$). From here on, all the reported error bars represent the $95\%$ confidence level (C.L.), unless otherwise stated. For reference we also show the constraint values for the $\Lambda$CDM model. 

We note that  GCCG prefers slightly higher central  values of $\sigma_8^0$ with respect to $\Lambda$CDM for both the combination of data sets, while the model with massive neutrinos prefers lower values. We note that assuming a $\Lambda$CDM scenario a tension at $3.2 \sigma$ in the estimation of $\sigma_8^0$ between Planck data  and KiDS+VIKING-450 combined with  DES-Y1 exists~\cite{Asgari:2019fkq}. In GCCG such tension is still present and to definitively settle the controversy,  the current analysis should be completed by using  datasets of weak lensing measurements, e.g. KiDS \cite{deJong:2015wca}. This  would require to consider non-linear effects in the MCMC analysis,   which is beyond the scope of this paper.  In the case of $\Omega_m^0$ we notice that  the $PBRS$ data sets makes its central value the same for both models. Planck data instead prefers lower values of $\Omega_m^0$ compared to $\Lambda$CDM. The inclusion of massive neutrinos slightly increases it central value in both models.   The bounds on the present day value of the Hubble function, $H_0$, in the case of $Planck$ alone are~:
\ba
&&H_0=  68 \pm 2  \,\, km\,s^{-1}Mpc^{-1}\,\,\, \mbox{for $\Lambda$CDM}\,, \\
&&H_0=  72^{+8}_{-5} \,\,km\,s^{-1}Mpc^{-1}\,\,\, \mbox{for GCCG}\,.
\ea
Direct measurements of $H_0$ at low redshift set its value to be $H_0 = 74.03\pm1.42$   km\,s$^{-1}$Mpc$^{-1}$~\cite{Riess:2019cxk}, whereas the results of the Planck Collaboration obtained by combining CMB data from the temperature and polarization maps and the lensing reconstruction, in the context of $\Lambda$CDM  favor lower values of $H_0$,  $H_0 = 67.4\pm0.5$   km\,s$^{-1}$Mpc$^{-1}$\cite{Aghanim:2018eyx} with a discrepancy which can reach the 4.4 $\sigma$~\cite{Riess:2019cxk}. In Fig.~\ref{fig:tension}, we  plot the two-dimensional observational contours at 68\% and 95\% C.L. for  $H_0$ and $\Omega_m^0$ constrained by the $Planck$  alone data (temperature and polarization) for both $\Lambda$CDM (solid black lines) and GCCG (red), we also include the low redshift measurement of $H_0$. From this Figure it is clear that although the bounds on $H_0$ for $\Lambda$CDM and GCCG are consistent with each other  within the errors, the GCCG model, unlike $\Lambda$CDM, is able to alleviate the tension of $H_0$ between the $Planck$ CMB data and its local measurements, which are compatible within $1\sigma$.   The constraint we found on $H_0$ is also fully compatible with the estimation obtained with the tip of the red giant branch  in the Large Magellanic Cloud, $H_0=72.4\pm 2$ km\,s$^{-1}$Mpc$^{-1}$ \cite{Yuan:2019npk}. The eased tension in the estimation of $H_0$ in the GCCG model is due to a difference in the late time background evolution, which is enhanced at low redshift, with respect to $\Lambda$CDM, as shown in Fig.~\ref{fig:hubble}. We note that if the tension between CMB data and low redshift measurements of $H_0$ disappears for the GCCG model, another tension arises which now is between the latter and BAO data as shown in Fig.~\ref{fig:tension} with yellow contours. In this case we note that BAO data assume a  fiducial flat $\Lambda$CDM cosmological model.  Although  BAO data can be used to constrain changes in the distance scale relative to that predicted by the $\Lambda$CDM model, the specific scenario we are investigating in this work involves a modification of the gravity force which might affect the result in a non-negligible way~\cite{Carter:2019ulk}. In this regard a further investigation is required.

Generally, we notice that the effect of massive neutrinos on the cosmological parameters of GCCG is to push their central values close to $\Lambda$CDM ones. This is due to the fact that they act in the direction to relieve the MG features as discussed in the previous section.   In Fig. \ref{fig:H0mnu} we show the marginalized 2-D joint distribution for  $H_0$ and $\sum m_\nu$ in both $\Lambda$CDM and GCCG. In the case of $\Lambda$CDM both datasets only set the upper bounds: $\sum m_\nu< 0.70$ eV for $Planck$ and  $\sum m_\nu< 0.23$ eV  for $PBRS$. For the GCCG model the $Planck$ data alone constrain the sum of neutrino masses to be $< 0.62$ eV. and also in this case the $PBRS$ data set a lower upper bound, which is   $\sum m_\nu< 0.45$ eV. We note that the full dataset is also able to detect a lower bound for the sum of neutrino masses which is 
\be
\Sigma m_\nu >0.11 \,\,\mbox{eV} \qquad \text{at} \, \,1 \,\sigma.
\ee

This result could be potentially interesting for present and future experiments which aim to find the absolute mass scale of neutrinos, such as KATRIN (see e.g.\ \cite{Aker:2019uuj}). We also note that this feature was already present in the covariant Galileon model~\cite{Peirone:2017vcq}.
 In Fig. \ref{fig:H0mnu},  we also note that the BAO+RSD+SNIa data in both cosmologies have the power to break the degeneracy between $H_0$ and $\sum m_\nu$.

Let us now discuss about  the cosmological constraints on the model parameters $q, s$. They are constrained to be strictly positive in agreement with the stability conditions. 
 The  data we use are able to constrain the parameter $s$ to be $0.6^{+1.7}_{-0.6}$. with $Planck$ alone, while $PBRS$ cuts the larger values so that  $s=0.05^{+0.08}_{-0.05}$.  The latter is due to the inclusion of BAO data which strongly constrain $s$ at background.  When massive neutrinos are included  the bound  for the complete data set is looser while it does not change in the case of $Planck$ alone.  The parameter $s$ shows  a degeneracy with $q$ as it can be seen in Fig. \ref{fig:sqcut}. Thus because $s$ is close to zero $q$ can span from 0 to very large values and as such it only shows  lower bounds. We also note another degeneracy between $s$ and $H_0$ parameters  (see right panel in Figure \ref{fig:degeneracysH0}). According to which  higher values of $H_0$ select higher values of $s$ and vice versa.  We then analyzed the constraints when on top of the $Planck$ data we also include the $H_0$ data point in~\cite{Riess:2019cxk}. The analysis with $Planck+H_0$ introduces a lower bound $s>0.13$. Because of the degeneracy between the two GCCG parameters, the latter translates into the upper bound $q<6.2$. We notice that this is the only dataset that sets a constraint on $q$, since even in the analysis with the full $PBRS$ this parameter is always unconstrained.
  
\begin{figure}[t!]
\centering
\includegraphics[width=0.4\textwidth]{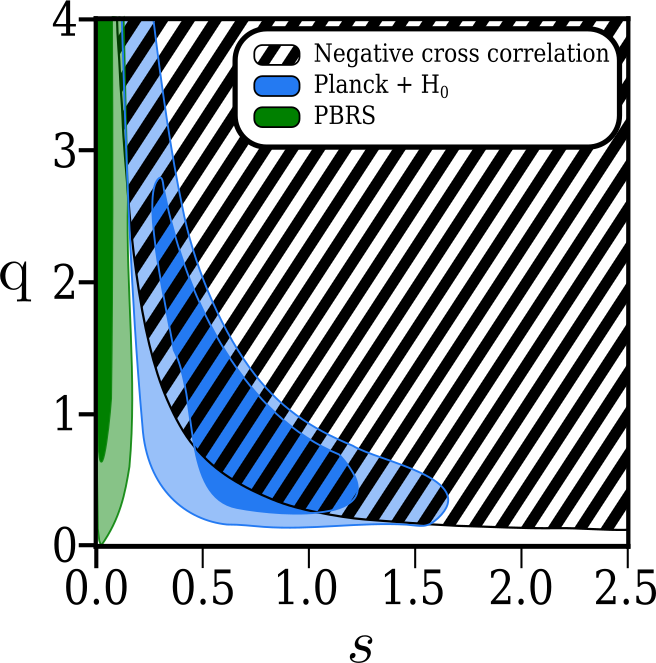}
\caption{\label{fig:isw}   Sign of the ISW-Galaxy cross correlation in the $\{s,q\}$-plane compared with the marginalized distributions obtained in the present work. The black dashed area represents the parameter space which is related to a negative sign of the ISW-Galaxy cross correlation. The latter has been calculated following the procedure in~\cite{Giacomello:2018jfi} and using the best fit values in Table~\ref{tab_bestfit_params}. The marginalized 2D distributions are plotted in blue for $Planck+H_0$ and in green for the $PBRS$ dataset. }
 \end{figure}

The  cross-correlation between the ISW signal and the matter (galaxy) distribution is known to be  a  powerful tool to test gravity \cite{Crittenden:1995ak,Boughn:1997vs,Kimura:2011td}. For the GCCG model, it has been identified  a viable region in the parameter space  that allows for positive ISW-Galaxy cross-correlation~\cite{Giacomello:2018jfi}. In this work  we use the methodology in~\cite{Giacomello:2018jfi} where it is assumed that only the model parameters $q$ and $s$ are free parameters and the other cosmological parameters are fixed. In our study we set the values of the cosmological parameters to the best fit values in Tab.~\ref{tab_bestfit_params}. In Figure~\ref{fig:isw} we show the results for the sign of the ISW-Galaxy cross-correlation in the $\{s,q\}$-plane where the black dashed area identifies the parameter space with negative cross-correlation. We overlap the marginalized 2-D joint distributions of the parameters $\{s, q\}$ for the $Planck+H_0$ and $PBRS$ data sets.  We note that the constraints obtained with the complete data set  lay in the region with a positive ISW-Galaxy cross-correlation, while for $Planck+H_0$ data the negative ISW-Galaxy cross-correlation cuts a part of the contours. We have verified that these results are independent of the chosen values for the cosmological parameters within their errors.
 
In order to quantify the preference of the GCCG model with respect to $\Lambda$CDM we make use of the Deviance Information Criterion (DIC)~\cite{RSSB:RSSB12062}:
\be
\text{DIC}:= \chi_\text{eff}^2 + 2 p_\text{D},
\ee
where $\chi_\text{eff}^2$ is the value of the effective $\chi^2$ corresponding to the maximum likelihood and $p_\text{D} = \overline{\chi}_\text{eff}^2 - \chi_\text{eff}^2$, here the bar indicates the average of the posterior distribution. The DIC accounts for both the goodness of fit ($\chi_\text{eff}^2$) and for the bayesian complexity of the model ($p_\text{D}$), disfavoring more complex models. The two cosmologies can then be compare by computing the following quantity
\be
\Delta \text{DIC} = \text{DIC}_\text{GCCG} - \text{DIC}_\text{$\Lambda$CDM}.
\ee
A negative $\Delta \text{DIC}$ supports the GCCG model over the $\Lambda$CDM one.
 In Tab.~\ref{tab_chi2} we show the values for both the $\Delta \chi_\text{eff}^2$ and  $\Delta \text{DIC}$, computed from the analyses with the $Planck$ and $PBRS$ datasets. We notice that, in both cases GCCG produces a lower $\chi_\text{eff}^2$ compared to $\Lambda$CDM: this is due to the fact that the model is able to lower the low-$\ell$ ISW tail of the CMB TT power spectrum, as shown in Figure~\ref{fig:observables} top left panel. In fact, when analyzing CMB data alone we see that the DIC favors the GCCG model ($\Delta \text{DIC} =-3.7$). Nevertheless, when considering the more exhaustive dataset $PBRS$ we found that the improvement in  $\chi_\text{eff}^2$ is not enough to compensate the increased model complexity of GCCG. In this case  $\Lambda$CDM becomes the preferred cosmology ($\Delta \text{DIC} =1.1$). We notice the same trend  in the analyses with massive neutrinos. 
 \renewcommand{\arraystretch}{1.4}
\begin{center}
\begin{table}[h!]
    \begin{tabular}{| l | c | c |}
    \hline
    	Dataset 	\,\,\,\,	& \,\, $\Delta \chi_\text{eff}^2$  \,\,    & \,\, $\Delta$DIC       \,\,		         \\ \hline  	        	
	$Planck$ 		&  $-4.9$                    & $-3.7$        		                 \\ \hline  	        
	$Planck$$+\nu$ 	&  $-6.5$                    & $-3.4$        		                 \\ \hline  	        
	$PBRS$ 		&  $-0.1$                    &  $1.1$        		                 \\ \hline  	    
  	$PBRS$$+\nu$	&  $-0.6$                    &  $1.1$        		                 \\ \hline  	    
     \end{tabular}
     \caption{\label{tab_chi2} Values of $\Delta \chi_\text{eff}^2$ and  $\Delta \text{DIC}$ computed between GCCG and $\Lambda$CDM for the $Planck$ and $PBRS$ datasets with and without massive neutrinos.   }
 \end{table}
\end{center}

We conclude this Section discussing the recent theoretical bound \cite{Creminelli:2019kjy} on the braiding function $\alpha_B$ which is characteristic of Horndeski models with  $G_3(X)\neq 0$.   GWs of sufficiently large amplitude produced by typical binary systems  might generate ghost and gradient instabilities in the dark energy perturbations for values of $|\alpha_B|\gsim10^{-2}$. This would lead to exclude models with cubic term such as  the one  considered in the present investigation  for which $\alpha_B=-\gamma_2aH_0/\hub$. We would like to notice that the  frequency considered in \cite{Creminelli:2019kjy} lies on the cut-off  of the EFT description. In this regards the EFT parameters could be dependent on the energy scale in  such a way that their values measured at low-energy scales may receive corrections when approaching larger frequency.  Indeed, the latter has been shown to be  the case  of the tensor speed of propagation \cite{deRham:2018red}. 
Furthermore, the results of \cite{Creminelli:2019kjy} seems to imply that we cannot neglect higher order perturbations terms compared to lower order ones.  It would be interesting to study if  such an instability exists in the context of a fully self-consistent second-order cosmological perturbation theory (i.e.\ considering all dynamical terms) and investigate how it depends on the parameters of the theory.
For all these reasons, we think more analytical work is needed to understand the possible influence of such a phenomenon. 
In the present work at low energy and present time we find  $\alpha_B^0>0$ for $Planck$ and $PBRS$. This is due to the degeneracy between $s$ and $q$ and to the fact that $q$ is unconstrained. When including the $H_0$ data point to $Planck$ we find the constraint  $\alpha_B^0=0.8\pm0.1$ at $1\,\sigma$ level. Such bound is one order of magnitude larger than the upper limit found in \cite{Creminelli:2019kjy}. Our constraint shows a completely independent bound on $\alpha_B^0$ based on cosmological data only.

%%%%%%%%%%%%%%%%%%%%%%%%%%%%%%%%%
\subsection{Impact of mass condition on cosmological constraints}
%%%%%%%%%%%%%%%%%%%%%%%%%%%%%%%%%

\begin{figure*}[t!]
\centering
\includegraphics[width=0.6\textwidth]{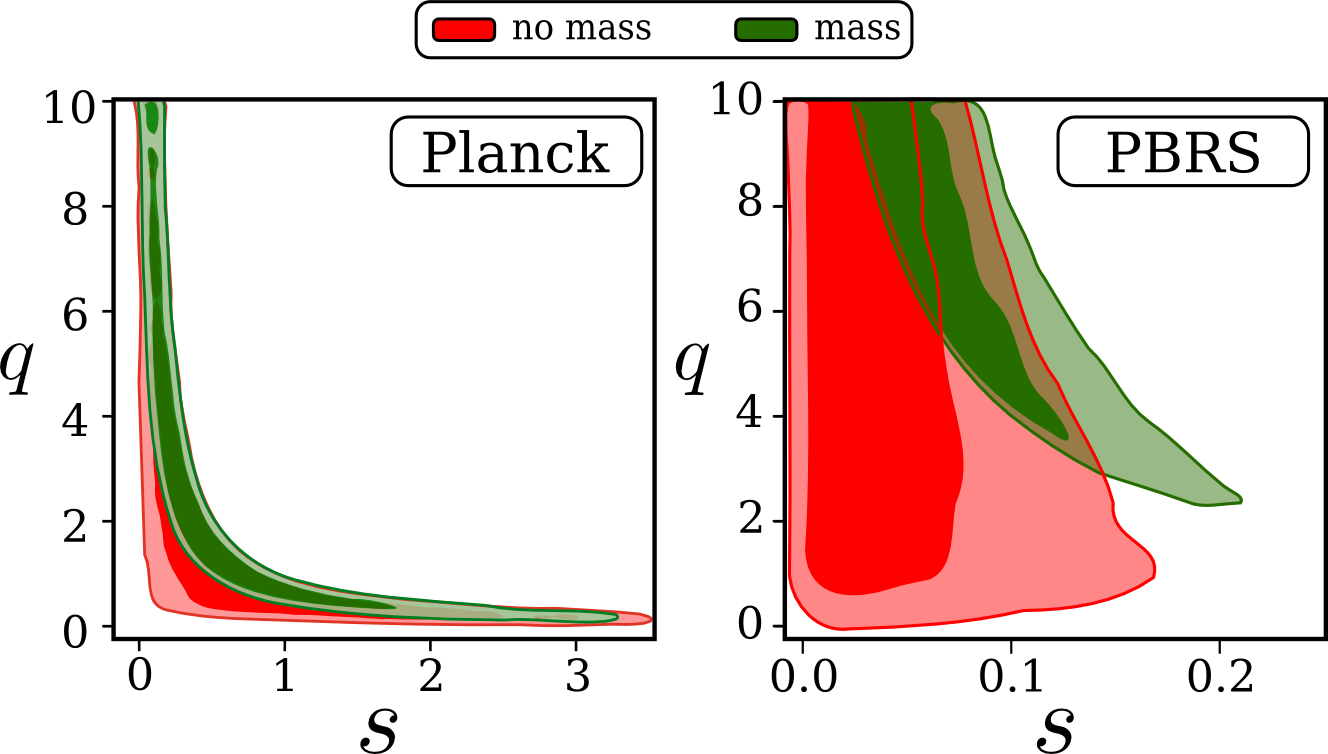}
\caption{\label{fig:sqcut}  The marginalized 2-D joint distribution for the model parameters $s$ and $q$ obtained from the analysis of the $Planck$ data. The inner region of the distribution represents the $68\%$ confidence level, while the outer region cuts the distribution at $95\%$. Different colors shows the effects of different stability cuts: in red we show the results when no-ghost, no-gradient  and strong coupling conditions are applied, in green the results of adding the mass stability condition.} 
 \end{figure*}

In Sec. \ref{Sec:stability} we discussed how the parameter space changes when including the mass condition as prior on top of the baseline stability conditions. In this section we discuss the impact of such condition on the cosmological and model parameters when $\beta=1$. 
We notice that  such condition does not affect the constraints on the cosmological parameters,  while it has an effect only on the model parameters. We show in Figure \ref{fig:sqcut},  the confidence regions for $q$ and $s$ obtained using the Planck  CMB temperature and polarization data when only the baseline stability conditions are  imposed (red) and the case in which the mass condition is included (green). As expected the constraints follow the shape of the stability cut induced by the mass condition.  As previously discussed,  the parameter space of $s$ and $q$ does not change when considering higher values of $\beta$ ($\beta\leq10$),  thus the green contours in Figure \ref{fig:sqcut} hold also for such values of $\beta$. From our analysis we note that $Planck$ alone does not add much information to the stability condition $q>0$, while the inclusion of BAO data in the complete dataset push  the lower bound to an higher value, $q>3.3$ for the model without massive neutrinos and $q>2$ with massive neutrinos.  The higher lower bound on $q$ obtained with the extended data sets is motivated by the fact that BAO+RSD+SNIa data (in particular BAO) strongly constrain the $s$ parameters at background level. For this case we obtain $s=0.44^{+1.5}_{-0.44}$ with $Planck$ alone, while $PBRS$ give the tighter constraint $s=0.082^{+0.083}_{-0.053}$. The marginalized constraint on $s$ with massive neutrinos does not change for the $PBRS$ data, while its central value with $Planck$ data alone is smaller $s=0.25^{+0.73}_{-0.25}$. Thus BAO data push the values of $s$ toward zero and as consequences they  select higher values of $q$. The latter is a consequence  of the  mass condition according to which small value of $s$ select higher values of $q$.

%-----------------------------------------------------------------------
 %-----------------------------------------------------------------------

%-----------------------------------------------------------------------

\renewcommand{\arraystretch}{1.3}
\begin{center}
\begin{table*}[t!]
    \begin{tabular}{| l | c | c | c | c | c | c |}
    \hline
    	Model 						&  $\sigma_8^0$                           &  $\Omega_m^0$       		                 & $H_0$                       & $q$                              & $s$                              & $\Sigma m_\nu$ (eV)      \\ \hline  \hline 
	$\Lambda$CDM ($Planck$)			& $  0.84 \pm 0.03  $ 	        & $    0.31 \pm 0.03  $ 	           	& $  68 \pm 2     $         &   -	                           & -                                   &  -    \\
	$\Lambda$CDM ($PBRS$)			& $  0.82\pm 0.03   $ 	        & $    0.30 \pm 0.01  $ 	           	& $  68.1 \pm 0.9     $         &   -	                           & -                                   &  -    \\  \hline  
	$\Lambda$CDM+$\nu$ ($Planck$)			& $  0.79^{+0.07}_{-0.10} $ 	& $   0.34^{+0.08}_{-0.05} $ 		& $  65.5^{+3.9}_{-5.3}    $    &   -	                           & -                      & $  <0.70   $  \\
	$\Lambda$CDM+$\nu$ ($PBRS$)			& $  0.81^{+0.03}_{-0.04} $ 	& $   0.31\pm 0.01 $ 		& $  67.79^{+0.95}_{-0.96}    $    &   -	                           & -                      & $  <0.23  $  \\ \hline  
	   \hline
	GCCG ($Planck$)		             	        & $  0.88^{+0.07}_{-0.05}  $ 	& $   0.27^{+0.05}_{-0.06}   $ 		& $  72^{+8}_{-5}     $   & $ >0 $	                   & $ 0.6^{+1.7}_{-0.6}$    &  -  \\
	GCCG ($PBRS$)			             	        & $  0.83\pm 0.03  $ 	& $   0.30\pm 0.01  $ 		& $  68.4\pm 0.9    $   & $ >0.8 $	& $ 0.05^{+0.08 }_{-0.05}$    &  -  \\ \hline 
	GCCG + $\nu$ ($Planck$)   	& $  0.8 \pm 0.1 $ 	& $   0.29\pm 0.08 $ 		& $  70^{+10}_{-8}    $    &  $ > 0 $	  &  $0.6^{+1.5}_{-0.6}      $                      & $  <0.62  $   \\
	GCCG + $\nu$  ($PBRS$)  	& $  0.79\pm0.04  $ 	         & $   0.31\pm 0.01    $ 		& $  68.0 \pm 1.1     $         & $ >1 $	  & $ 0.1^{+0.2}_{-0.1}$                      & $ <0.45 $   \\
	    \hline
     \end{tabular}
     \caption{\label{tab_bestfit_params} Marginalized constraints on cosmological and model parameters at $95\%$ confidence level. For the GCCG model the  baseline stability conditions (no-ghost, no-gradient and strong coupling conditions)  are always assumed.   }
 \end{table*}
\end{center}
\renewcommand{\arraystretch}{1}
%-----------------------------------------------------------------------

\section{Conclusion}\label{Sec:conclusion}

In this work we  have performed a thorough investigation of the impact  on the cosmological observables of the modification induced by a specific class of Galileon models, the generalized cubic covariant Galileon (GCCG).  Compared to $\Lambda$CDM  the model  introduces two extra parameters $\{s,q\}$ of which only $s$ affects the background dynamics, while both introduce modifications at linear perturbation level.  We have identified modifications in the ISW effect, the gravitational lensing, the rate of growth of structure and  temperature-temperature power spectrum at large angular scales. While both the lensing and the matter power spectra are in general enhanced with respect to  $\Lambda$CDM, the low-$\ell$ TT power spectrum can be either enhanced or suppressed and the high-$\ell$ peaks are shifted due to a modified background evolution.  We found that the inclusion of massive neutrinos generally brings down the deviations of GCCG  from $\Lambda$CDM. We have performed cosmological constraints considering data from  CMB, BAO, RSD, SNIa and Cepheids ($H_0$)  in different combinations in order to identify  how different data sets contribute to the parameter bounds. The results are shown in Tab. \ref{tab_bestfit_params}. Notably, we found that GCCG is able to ease the tension in the estimation of the present day value of the Hubble parameter $H_0$ between Planck and low-$z$ measurements which arises within $\Lambda$CDM. However the tension is again present when we include BAO data. This case needs a further investigation as BAO data might be biased toward $\Lambda$CDM-like models~\cite{Carter:2019ulk}. 
We found that the tightest constraints for $s=0.05^{+0.08 }_{-0.05}$ at 95\% C.L are for the $PBRS$ data set, while $q$ shows only a lower bound $q>0.8$. The joint analysis with Planck and $H_0$ instead is able to set the upper bound $q<0.62$.  We also found a lower bound for the sum of the neutrino masses to be $>0.11$ eV at $1\sigma$ along with the usual upper bound. This is an interesting result which should be further considered in light of  future results from experiments measuring the neutrino mass.
 
The model selection analysis shows that  the extended model is favored over \lcdm when considering the Planck data alone, because the GCCG is able to better fit the ISW tail. Nevertheless, the complete dataset points toward the standard cosmological model with a DIC value that indicates a mild preference for the latter. These results suggest that further work is needed in order to asses the statistical preference of one cosmological model over the other and it will be the subject of upcoming projects.

This study is a further proof that Galileon models cannot be definitely excluded with respect to \lcdm. Indeed besides the model investigated here there is another case in which the data support the Galileon cosmology over \lcdm, the Galileon ghost condensate model \cite{Peirone:2019aua}. In this regard it will be essential to test Galileon models with next generation surveys, which will offer us the possibility to test gravity at cosmological scales with unprecedented accuracy.

\acknowledgments

We thank M.\ Raveri, A.\ Silvestri, M.\ Vicinanza, and L.\ Yin for useful discussions.
NF and LA are  supported by Funda\c{c}\~{a}o para a  Ci\^{e}ncia e a Tecnologia (FCT) through national funds  (UID/FIS/04434/2019), by FEDER through COMPETE2020  (POCI-01-0145-FEDER-007672). The research of NF is also supported by FCT project ``DarkRipple -- Spacetime ripples in the dark gravitational Universe" with ref.~number PTDC/FIS-OUT/29048/2017.
SP acknowledges support from the NWO and the Dutch Ministry of Education, Culture and Science (OCW), and also from the D-ITP consortium, a program of the NWO that is funded by the OCW.
NF and  SP  acknowledge the COST Action  (CANTATA/CA15117), supported by COST (European Cooperation in  Science and Technology).

\bibliography{GCG}

%merlin.mbs apsrev4-1.bst 2010-07-25 4.21a (PWD, AO, DPC) hacked
%Control: key (0)
%Control: author (8) initials jnrlst
%Control: editor formatted (1) identically to author
%Control: production of article title (-1) disabled
%Control: page (0) single
%Control: year (1) truncated
%Control: production of eprint (0) enabled
\begin{thebibliography}{99}%
\makeatletter
\providecommand \@ifxundefined [1]{%
 \@ifx{#1\undefined}
}%
\providecommand \@ifnum [1]{%
 \ifnum #1\expandafter \@firstoftwo
 \else \expandafter \@secondoftwo
 \fi
}%
\providecommand \@ifx [1]{%
 \ifx #1\expandafter \@firstoftwo
 \else \expandafter \@secondoftwo
 \fi
}%
\providecommand \natexlab [1]{#1}%
\providecommand \enquote  [1]{``#1''}%
\providecommand \bibnamefont  [1]{#1}%
\providecommand \bibfnamefont [1]{#1}%
\providecommand \citenamefont [1]{#1}%
\providecommand \href@noop [0]{\@secondoftwo}%
\providecommand \href [0]{\begingroup \@sanitize@url \@href}%
\providecommand \@href[1]{\@@startlink{#1}\@@href}%
\providecommand \@@href[1]{\endgroup#1\@@endlink}%
\providecommand \@sanitize@url [0]{\catcode `\\12\catcode `\$12\catcode
  `\&12\catcode `\#12\catcode `\^12\catcode `\_12\catcode `\%12\relax}%
\providecommand \@@startlink[1]{}%
\providecommand \@@endlink[0]{}%
\providecommand \url  [0]{\begingroup\@sanitize@url \@url }%
\providecommand \@url [1]{\endgroup\@href {#1}{\urlprefix }}%
\providecommand \urlprefix  [0]{URL }%
\providecommand \Eprint [0]{\href }%
\providecommand \doibase [0]{http://dx.doi.org/}%
\providecommand \selectlanguage [0]{\@gobble}%
\providecommand \bibinfo  [0]{\@secondoftwo}%
\providecommand \bibfield  [0]{\@secondoftwo}%
\providecommand \translation [1]{[#1]}%
\providecommand \BibitemOpen [0]{}%
\providecommand \bibitemStop [0]{}%
\providecommand \bibitemNoStop [0]{.\EOS\space}%
\providecommand \EOS [0]{\spacefactor3000\relax}%
\providecommand \BibitemShut  [1]{\csname bibitem#1\endcsname}%
\let\auto@bib@innerbib\@empty
%</preamble>
\bibitem [{\citenamefont {Joyce}\ \emph {et~al.}(2015)\citenamefont {Joyce},
  \citenamefont {Jain}, \citenamefont {Khoury},\ and\ \citenamefont
  {Trodden}}]{Joyce:2014kja}%
  \BibitemOpen
  \bibfield  {author} {\bibinfo {author} {\bibfnamefont {A.}~\bibnamefont
  {Joyce}}, \bibinfo {author} {\bibfnamefont {B.}~\bibnamefont {Jain}},
  \bibinfo {author} {\bibfnamefont {J.}~\bibnamefont {Khoury}}, \ and\ \bibinfo
  {author} {\bibfnamefont {M.}~\bibnamefont {Trodden}},\ }\href {\doibase
  10.1016/j.physrep.2014.12.002} {\bibfield  {journal} {\bibinfo  {journal}
  {Phys. Rept.}\ }\textbf {\bibinfo {volume} {568}},\ \bibinfo {pages} {1}
  (\bibinfo {year} {2015})},\ \Eprint {http://arxiv.org/abs/1407.0059}
  {arXiv:1407.0059 [astro-ph.CO]} \BibitemShut {NoStop}%
%%CITATION = ARXIV:1407.0059;%%
\bibitem [{\citenamefont {Lue}\ \emph {et~al.}(2004)\citenamefont {Lue},
  \citenamefont {Scoccimarro},\ and\ \citenamefont {Starkman}}]{Lue:2004rj}%
  \BibitemOpen
  \bibfield  {author} {\bibinfo {author} {\bibfnamefont {A.}~\bibnamefont
  {Lue}}, \bibinfo {author} {\bibfnamefont {R.}~\bibnamefont {Scoccimarro}}, \
  and\ \bibinfo {author} {\bibfnamefont {G.~D.}\ \bibnamefont {Starkman}},\
  }\href {\doibase 10.1103/PhysRevD.69.124015} {\bibfield  {journal} {\bibinfo
  {journal} {Phys. Rev.}\ }\textbf {\bibinfo {volume} {D69}},\ \bibinfo {pages}
  {124015} (\bibinfo {year} {2004})},\ \Eprint
  {http://arxiv.org/abs/astro-ph/0401515} {arXiv:astro-ph/0401515 [astro-ph]}
  \BibitemShut {NoStop}%
%%CITATION = ASTRO-PH/0401515;%%
\bibitem [{\citenamefont {Copeland}\ \emph {et~al.}(2006)\citenamefont
  {Copeland}, \citenamefont {Sami},\ and\ \citenamefont
  {Tsujikawa}}]{Copeland:2006wr}%
  \BibitemOpen
  \bibfield  {author} {\bibinfo {author} {\bibfnamefont {E.~J.}\ \bibnamefont
  {Copeland}}, \bibinfo {author} {\bibfnamefont {M.}~\bibnamefont {Sami}}, \
  and\ \bibinfo {author} {\bibfnamefont {S.}~\bibnamefont {Tsujikawa}},\ }\href
  {\doibase 10.1142/S021827180600942X} {\bibfield  {journal} {\bibinfo
  {journal} {Int. J. Mod. Phys.}\ }\textbf {\bibinfo {volume} {D15}},\ \bibinfo
  {pages} {1753} (\bibinfo {year} {2006})},\ \Eprint
  {http://arxiv.org/abs/hep-th/0603057} {arXiv:hep-th/0603057 [hep-th]}
  \BibitemShut {NoStop}%
%%CITATION = HEP-TH/0603057;%%
\bibitem [{\citenamefont {Silvestri}\ and\ \citenamefont
  {Trodden}(2009)}]{Silvestri:2009hh}%
  \BibitemOpen
  \bibfield  {author} {\bibinfo {author} {\bibfnamefont {A.}~\bibnamefont
  {Silvestri}}\ and\ \bibinfo {author} {\bibfnamefont {M.}~\bibnamefont
  {Trodden}},\ }\href {\doibase 10.1088/0034-4885/72/9/096901} {\bibfield
  {journal} {\bibinfo  {journal} {Rept. Prog. Phys.}\ }\textbf {\bibinfo
  {volume} {72}},\ \bibinfo {pages} {096901} (\bibinfo {year} {2009})},\
  \Eprint {http://arxiv.org/abs/0904.0024} {arXiv:0904.0024 [astro-ph.CO]}
  \BibitemShut {NoStop}%
%%CITATION = ARXIV:0904.0024;%%
\bibitem [{\citenamefont {Capozziello}\ and\ \citenamefont
  {De~Laurentis}(2011)}]{Capozziello:2011et}%
  \BibitemOpen
  \bibfield  {author} {\bibinfo {author} {\bibfnamefont {S.}~\bibnamefont
  {Capozziello}}\ and\ \bibinfo {author} {\bibfnamefont {M.}~\bibnamefont
  {De~Laurentis}},\ }\href {\doibase 10.1016/j.physrep.2011.09.003} {\bibfield
  {journal} {\bibinfo  {journal} {Phys. Rept.}\ }\textbf {\bibinfo {volume}
  {509}},\ \bibinfo {pages} {167} (\bibinfo {year} {2011})},\ \Eprint
  {http://arxiv.org/abs/1108.6266} {arXiv:1108.6266 [gr-qc]} \BibitemShut
  {NoStop}%
%%CITATION = ARXIV:1108.6266;%%
\bibitem [{\citenamefont {Clifton}\ \emph {et~al.}(2012)\citenamefont
  {Clifton}, \citenamefont {Ferreira}, \citenamefont {Padilla},\ and\
  \citenamefont {Skordis}}]{Clifton:2011jh}%
  \BibitemOpen
  \bibfield  {author} {\bibinfo {author} {\bibfnamefont {T.}~\bibnamefont
  {Clifton}}, \bibinfo {author} {\bibfnamefont {P.~G.}\ \bibnamefont
  {Ferreira}}, \bibinfo {author} {\bibfnamefont {A.}~\bibnamefont {Padilla}}, \
  and\ \bibinfo {author} {\bibfnamefont {C.}~\bibnamefont {Skordis}},\ }\href
  {\doibase 10.1016/j.physrep.2012.01.001} {\bibfield  {journal} {\bibinfo
  {journal} {Phys. Rept.}\ }\textbf {\bibinfo {volume} {513}},\ \bibinfo
  {pages} {1} (\bibinfo {year} {2012})},\ \Eprint
  {http://arxiv.org/abs/1106.2476} {arXiv:1106.2476 [astro-ph.CO]} \BibitemShut
  {NoStop}%
%%CITATION = ARXIV:1106.2476;%%
\bibitem [{\citenamefont {Tsujikawa}(2010)}]{Tsujikawa:2010zza}%
  \BibitemOpen
  \bibfield  {author} {\bibinfo {author} {\bibfnamefont {S.}~\bibnamefont
  {Tsujikawa}},\ }\href {\doibase 10.1007/978-3-642-10598-2_3} {\bibfield
  {journal} {\bibinfo  {journal} {Lect. Notes Phys.}\ }\textbf {\bibinfo
  {volume} {800}},\ \bibinfo {pages} {99} (\bibinfo {year} {2010})},\ \Eprint
  {http://arxiv.org/abs/1101.0191} {arXiv:1101.0191 [gr-qc]} \BibitemShut
  {NoStop}%
%%CITATION = ARXIV:1101.0191;%%
\bibitem [{\citenamefont {Koyama}(2016)}]{Koyama:2015vza}%
  \BibitemOpen
  \bibfield  {author} {\bibinfo {author} {\bibfnamefont {K.}~\bibnamefont
  {Koyama}},\ }\href {\doibase 10.1088/0034-4885/79/4/046902} {\bibfield
  {journal} {\bibinfo  {journal} {Rept. Prog. Phys.}\ }\textbf {\bibinfo
  {volume} {79}},\ \bibinfo {pages} {046902} (\bibinfo {year} {2016})},\
  \Eprint {http://arxiv.org/abs/1504.04623} {arXiv:1504.04623 [astro-ph.CO]}
  \BibitemShut {NoStop}%
%%CITATION = ARXIV:1504.04623;%%
\bibitem [{\citenamefont {Ferreira}(2019)}]{Ferreira:2019xrr}%
  \BibitemOpen
  \bibfield  {author} {\bibinfo {author} {\bibfnamefont {P.~G.}\ \bibnamefont
  {Ferreira}},\ }\href {\doibase 10.1146/annurev-astro-091918-104423} {\
  (\bibinfo {year} {2019}),\ 10.1146/annurev-astro-091918-104423},\ \Eprint
  {http://arxiv.org/abs/1902.10503} {arXiv:1902.10503 [astro-ph.CO]}
  \BibitemShut {NoStop}%
%%CITATION = ARXIV:1902.10503;%%
\bibitem [{\citenamefont {Kobayashi}(2019)}]{Kobayashi:2019hrl}%
  \BibitemOpen
  \bibfield  {author} {\bibinfo {author} {\bibfnamefont {T.}~\bibnamefont
  {Kobayashi}},\ }\href {\doibase 10.1088/1361-6633/ab2429} {\bibfield
  {journal} {\bibinfo  {journal} {Rept. Prog. Phys.}\ }\textbf {\bibinfo
  {volume} {82}},\ \bibinfo {pages} {086901} (\bibinfo {year} {2019})},\
  \Eprint {http://arxiv.org/abs/1901.07183} {arXiv:1901.07183 [gr-qc]}
  \BibitemShut {NoStop}%
%%CITATION = ARXIV:1901.07183;%%
\bibitem [{\citenamefont {Horndeski}(1974)}]{Horndeski:1974wa}%
  \BibitemOpen
  \bibfield  {author} {\bibinfo {author} {\bibfnamefont {G.~W.}\ \bibnamefont
  {Horndeski}},\ }\href {\doibase 10.1007/BF01807638} {\bibfield  {journal}
  {\bibinfo  {journal} {Int. J. Theor. Phys.}\ }\textbf {\bibinfo {volume}
  {10}},\ \bibinfo {pages} {363} (\bibinfo {year} {1974})}\BibitemShut
  {NoStop}%
%%CITATION = IJTPB,10,363;%%
\bibitem [{\citenamefont {Deffayet}\ \emph
  {et~al.}(2009{\natexlab{a}})\citenamefont {Deffayet}, \citenamefont {Deser},\
  and\ \citenamefont {Esposito-Farese}}]{Deffayet:2009mn}%
  \BibitemOpen
  \bibfield  {author} {\bibinfo {author} {\bibfnamefont {C.}~\bibnamefont
  {Deffayet}}, \bibinfo {author} {\bibfnamefont {S.}~\bibnamefont {Deser}}, \
  and\ \bibinfo {author} {\bibfnamefont {G.}~\bibnamefont {Esposito-Farese}},\
  }\href {\doibase 10.1103/PhysRevD.80.064015} {\bibfield  {journal} {\bibinfo
  {journal} {Phys. Rev.}\ }\textbf {\bibinfo {volume} {D80}},\ \bibinfo {pages}
  {064015} (\bibinfo {year} {2009}{\natexlab{a}})},\ \Eprint
  {http://arxiv.org/abs/0906.1967} {arXiv:0906.1967 [gr-qc]} \BibitemShut
  {NoStop}%
%%CITATION = ARXIV:0906.1967;%%
\bibitem [{\citenamefont {Kase}\ and\ \citenamefont
  {Tsujikawa}(2018)}]{Kase:2018iwp}%
  \BibitemOpen
  \bibfield  {author} {\bibinfo {author} {\bibfnamefont {R.}~\bibnamefont
  {Kase}}\ and\ \bibinfo {author} {\bibfnamefont {S.}~\bibnamefont
  {Tsujikawa}},\ }\href {\doibase 10.1103/PhysRevD.97.103501} {\bibfield
  {journal} {\bibinfo  {journal} {Phys. Rev.}\ }\textbf {\bibinfo {volume}
  {D97}},\ \bibinfo {pages} {103501} (\bibinfo {year} {2018})},\ \Eprint
  {http://arxiv.org/abs/1802.02728} {arXiv:1802.02728 [gr-qc]} \BibitemShut
  {NoStop}%
%%CITATION = ARXIV:1802.02728;%%
\bibitem [{\citenamefont {Peirone}\ \emph {et~al.}(2019)\citenamefont
  {Peirone}, \citenamefont {Benevento}, \citenamefont {Frusciante},\ and\
  \citenamefont {Tsujikawa}}]{Peirone:2019aua}%
  \BibitemOpen
  \bibfield  {author} {\bibinfo {author} {\bibfnamefont {S.}~\bibnamefont
  {Peirone}}, \bibinfo {author} {\bibfnamefont {G.}~\bibnamefont {Benevento}},
  \bibinfo {author} {\bibfnamefont {N.}~\bibnamefont {Frusciante}}, \ and\
  \bibinfo {author} {\bibfnamefont {S.}~\bibnamefont {Tsujikawa}},\ }\href
  {\doibase 10.1103/PhysRevD.100.063540} {\bibfield  {journal} {\bibinfo
  {journal} {Phys. Rev.}\ }\textbf {\bibinfo {volume} {D100}},\ \bibinfo
  {pages} {063540} (\bibinfo {year} {2019})},\ \Eprint
  {http://arxiv.org/abs/1905.05166} {arXiv:1905.05166 [astro-ph.CO]}
  \BibitemShut {NoStop}%
%%CITATION = ARXIV:1905.05166;%%
\bibitem [{\citenamefont {De~Felice}\ and\ \citenamefont
  {Tsujikawa}(2012{\natexlab{a}})}]{DeFelice:2011bh}%
  \BibitemOpen
  \bibfield  {author} {\bibinfo {author} {\bibfnamefont {A.}~\bibnamefont
  {De~Felice}}\ and\ \bibinfo {author} {\bibfnamefont {S.}~\bibnamefont
  {Tsujikawa}},\ }\href {\doibase 10.1088/1475-7516/2012/02/007} {\bibfield
  {journal} {\bibinfo  {journal} {JCAP}\ }\textbf {\bibinfo {volume} {1202}},\
  \bibinfo {pages} {007} (\bibinfo {year} {2012}{\natexlab{a}})},\ \Eprint
  {http://arxiv.org/abs/1110.3878} {arXiv:1110.3878 [gr-qc]} \BibitemShut
  {NoStop}%
%%CITATION = ARXIV:1110.3878;%%
\bibitem [{\citenamefont {Deffayet}\ \emph
  {et~al.}(2009{\natexlab{b}})\citenamefont {Deffayet}, \citenamefont
  {Esposito-Farese},\ and\ \citenamefont {Vikman}}]{Deffayet:2009wt}%
  \BibitemOpen
  \bibfield  {author} {\bibinfo {author} {\bibfnamefont {C.}~\bibnamefont
  {Deffayet}}, \bibinfo {author} {\bibfnamefont {G.}~\bibnamefont
  {Esposito-Farese}}, \ and\ \bibinfo {author} {\bibfnamefont {A.}~\bibnamefont
  {Vikman}},\ }\href {\doibase 10.1103/PhysRevD.79.084003} {\bibfield
  {journal} {\bibinfo  {journal} {Phys. Rev.}\ }\textbf {\bibinfo {volume}
  {D79}},\ \bibinfo {pages} {084003} (\bibinfo {year} {2009}{\natexlab{b}})},\
  \Eprint {http://arxiv.org/abs/0901.1314} {arXiv:0901.1314 [hep-th]}
  \BibitemShut {NoStop}%
%%CITATION = ARXIV:0901.1314;%%
\bibitem [{\citenamefont {De~Felice}\ and\ \citenamefont
  {Tsujikawa}(2010)}]{DeFelice:2010pv}%
  \BibitemOpen
  \bibfield  {author} {\bibinfo {author} {\bibfnamefont {A.}~\bibnamefont
  {De~Felice}}\ and\ \bibinfo {author} {\bibfnamefont {S.}~\bibnamefont
  {Tsujikawa}},\ }\href {\doibase 10.1103/PhysRevLett.105.111301} {\bibfield
  {journal} {\bibinfo  {journal} {Phys. Rev. Lett.}\ }\textbf {\bibinfo
  {volume} {105}},\ \bibinfo {pages} {111301} (\bibinfo {year} {2010})},\
  \Eprint {http://arxiv.org/abs/1007.2700} {arXiv:1007.2700 [astro-ph.CO]}
  \BibitemShut {NoStop}%
%%CITATION = ARXIV:1007.2700;%%
\bibitem [{\citenamefont {De~Felice}\ and\ \citenamefont
  {Tsujikawa}(2011)}]{DeFelice:2010nf}%
  \BibitemOpen
  \bibfield  {author} {\bibinfo {author} {\bibfnamefont {A.}~\bibnamefont
  {De~Felice}}\ and\ \bibinfo {author} {\bibfnamefont {S.}~\bibnamefont
  {Tsujikawa}},\ }\href {\doibase 10.1103/PhysRevD.84.124029} {\bibfield
  {journal} {\bibinfo  {journal} {Phys. Rev.}\ }\textbf {\bibinfo {volume}
  {D84}},\ \bibinfo {pages} {124029} (\bibinfo {year} {2011})},\ \Eprint
  {http://arxiv.org/abs/1008.4236} {arXiv:1008.4236 [hep-th]} \BibitemShut
  {NoStop}%
%%CITATION = ARXIV:1008.4236;%%
\bibitem [{\citenamefont {Frusciante}\ \emph
  {et~al.}(2019{\natexlab{a}})\citenamefont {Frusciante}, \citenamefont {Kase},
  \citenamefont {Koyama}, \citenamefont {Tsujikawa},\ and\ \citenamefont
  {Vernieri}}]{Frusciante:2018tvu}%
  \BibitemOpen
  \bibfield  {author} {\bibinfo {author} {\bibfnamefont {N.}~\bibnamefont
  {Frusciante}}, \bibinfo {author} {\bibfnamefont {R.}~\bibnamefont {Kase}},
  \bibinfo {author} {\bibfnamefont {K.}~\bibnamefont {Koyama}}, \bibinfo
  {author} {\bibfnamefont {S.}~\bibnamefont {Tsujikawa}}, \ and\ \bibinfo
  {author} {\bibfnamefont {D.}~\bibnamefont {Vernieri}},\ }\href {\doibase
  10.1016/j.physletb.2019.01.009} {\bibfield  {journal} {\bibinfo  {journal}
  {Phys. Lett.}\ }\textbf {\bibinfo {volume} {B790}},\ \bibinfo {pages} {167}
  (\bibinfo {year} {2019}{\natexlab{a}})},\ \Eprint
  {http://arxiv.org/abs/1812.05204} {arXiv:1812.05204 [gr-qc]} \BibitemShut
  {NoStop}%
%%CITATION = ARXIV:1812.05204;%%
\bibitem [{\citenamefont {De~Felice}\ and\ \citenamefont
  {Tsujikawa}(2012{\natexlab{b}})}]{DeFelice:2011aa}%
  \BibitemOpen
  \bibfield  {author} {\bibinfo {author} {\bibfnamefont {A.}~\bibnamefont
  {De~Felice}}\ and\ \bibinfo {author} {\bibfnamefont {S.}~\bibnamefont
  {Tsujikawa}},\ }\href {\doibase 10.1088/1475-7516/2012/03/025} {\bibfield
  {journal} {\bibinfo  {journal} {JCAP}\ }\textbf {\bibinfo {volume} {1203}},\
  \bibinfo {pages} {025} (\bibinfo {year} {2012}{\natexlab{b}})},\ \Eprint
  {http://arxiv.org/abs/1112.1774} {arXiv:1112.1774 [astro-ph.CO]} \BibitemShut
  {NoStop}%
%%CITATION = ARXIV:1112.1774;%%
\bibitem [{\citenamefont {Renk}\ \emph {et~al.}(2017)\citenamefont {Renk},
  \citenamefont {Zumalacarregui}, \citenamefont {Montanari},\ and\
  \citenamefont {Barreira}}]{Renk:2017rzu}%
  \BibitemOpen
  \bibfield  {author} {\bibinfo {author} {\bibfnamefont {J.}~\bibnamefont
  {Renk}}, \bibinfo {author} {\bibfnamefont {M.}~\bibnamefont
  {Zumalacarregui}}, \bibinfo {author} {\bibfnamefont {F.}~\bibnamefont
  {Montanari}}, \ and\ \bibinfo {author} {\bibfnamefont {A.}~\bibnamefont
  {Barreira}},\ }\href {\doibase 10.1088/1475-7516/2017/10/020} {\bibfield
  {journal} {\bibinfo  {journal} {JCAP}\ }\textbf {\bibinfo {volume} {1710}},\
  \bibinfo {pages} {020} (\bibinfo {year} {2017})},\ \Eprint
  {http://arxiv.org/abs/1707.02263} {arXiv:1707.02263 [astro-ph.CO]}
  \BibitemShut {NoStop}%
%%CITATION = ARXIV:1707.02263;%%
\bibitem [{\citenamefont {Peirone}\ \emph {et~al.}(2018)\citenamefont
  {Peirone}, \citenamefont {Frusciante}, \citenamefont {Hu}, \citenamefont
  {Raveri},\ and\ \citenamefont {Silvestri}}]{Peirone:2017vcq}%
  \BibitemOpen
  \bibfield  {author} {\bibinfo {author} {\bibfnamefont {S.}~\bibnamefont
  {Peirone}}, \bibinfo {author} {\bibfnamefont {N.}~\bibnamefont {Frusciante}},
  \bibinfo {author} {\bibfnamefont {B.}~\bibnamefont {Hu}}, \bibinfo {author}
  {\bibfnamefont {M.}~\bibnamefont {Raveri}}, \ and\ \bibinfo {author}
  {\bibfnamefont {A.}~\bibnamefont {Silvestri}},\ }\href {\doibase
  10.1103/PhysRevD.97.063518} {\bibfield  {journal} {\bibinfo  {journal} {Phys.
  Rev.}\ }\textbf {\bibinfo {volume} {D97}},\ \bibinfo {pages} {063518}
  (\bibinfo {year} {2018})},\ \Eprint {http://arxiv.org/abs/1711.04760}
  {arXiv:1711.04760 [astro-ph.CO]} \BibitemShut {NoStop}%
%%CITATION = ARXIV:1711.04760;%%
\bibitem [{\citenamefont {Leloup}\ \emph {et~al.}(2019)\citenamefont {Leloup},
  \citenamefont {Ruhlmann-Kleider}, \citenamefont {Neveu},\ and\ \citenamefont
  {De~Mattia}}]{Leloup:2019fas}%
  \BibitemOpen
  \bibfield  {author} {\bibinfo {author} {\bibfnamefont {C.}~\bibnamefont
  {Leloup}}, \bibinfo {author} {\bibfnamefont {V.}~\bibnamefont
  {Ruhlmann-Kleider}}, \bibinfo {author} {\bibfnamefont {J.}~\bibnamefont
  {Neveu}}, \ and\ \bibinfo {author} {\bibfnamefont {A.}~\bibnamefont
  {De~Mattia}},\ }\href {\doibase 10.1088/1475-7516/2019/05/011} {\bibfield
  {journal} {\bibinfo  {journal} {JCAP}\ }\textbf {\bibinfo {volume} {1905}},\
  \bibinfo {pages} {011} (\bibinfo {year} {2019})},\ \Eprint
  {http://arxiv.org/abs/1902.07065} {arXiv:1902.07065 [astro-ph.CO]}
  \BibitemShut {NoStop}%
%%CITATION = ARXIV:1902.07065;%%
\bibitem [{\citenamefont {Abbott}\ \emph
  {et~al.}(2017{\natexlab{a}})\citenamefont {Abbott} \emph
  {et~al.}}]{TheLIGOScientific:2017qsa}%
  \BibitemOpen
  \bibfield  {author} {\bibinfo {author} {\bibfnamefont {B.}~\bibnamefont
  {Abbott}} \emph {et~al.} (\bibinfo {collaboration} {Virgo, LIGO
  Scientific}),\ }\href {\doibase 10.1103/PhysRevLett.119.161101} {\bibfield
  {journal} {\bibinfo  {journal} {Phys. Rev. Lett.}\ }\textbf {\bibinfo
  {volume} {119}},\ \bibinfo {pages} {161101} (\bibinfo {year}
  {2017}{\natexlab{a}})},\ \Eprint {http://arxiv.org/abs/1710.05832}
  {arXiv:1710.05832 [gr-qc]} \BibitemShut {NoStop}%
%%CITATION = ARXIV:1710.05832;%%
\bibitem [{\citenamefont {Coulter}\ \emph {et~al.}(2017)\citenamefont {Coulter}
  \emph {et~al.}}]{Coulter:2017wya}%
  \BibitemOpen
  \bibfield  {author} {\bibinfo {author} {\bibfnamefont {D.~A.}\ \bibnamefont
  {Coulter}} \emph {et~al.},\ }\href {\doibase 10.1126/science.aap9811}
  {\bibfield  {journal} {\bibinfo  {journal} {Science}\ } (\bibinfo {year}
  {2017}),\ 10.1126/science.aap9811},\ \Eprint
  {http://arxiv.org/abs/1710.05452} {arXiv:1710.05452 [astro-ph.HE]}
  \BibitemShut {NoStop}%
%%CITATION = ARXIV:1710.05452;%%
\bibitem [{\citenamefont {Abbott}\ \emph
  {et~al.}(2017{\natexlab{b}})\citenamefont {Abbott} \emph
  {et~al.}}]{GBM:2017lvd}%
  \BibitemOpen
  \bibfield  {author} {\bibinfo {author} {\bibfnamefont {B.~P.}\ \bibnamefont
  {Abbott}} \emph {et~al.},\ }\href {\doibase 10.3847/2041-8213/aa91c9}
  {\bibfield  {journal} {\bibinfo  {journal} {Astrophys. J.}\ }\textbf
  {\bibinfo {volume} {848}},\ \bibinfo {pages} {L12} (\bibinfo {year}
  {2017}{\natexlab{b}})},\ \Eprint {http://arxiv.org/abs/1710.05833}
  {arXiv:1710.05833 [astro-ph.HE]} \BibitemShut {NoStop}%
%%CITATION = ARXIV:1710.05833;%%
\bibitem [{\citenamefont {Creminelli}\ and\ \citenamefont
  {Vernizzi}(2017)}]{Creminelli:2017sry}%
  \BibitemOpen
  \bibfield  {author} {\bibinfo {author} {\bibfnamefont {P.}~\bibnamefont
  {Creminelli}}\ and\ \bibinfo {author} {\bibfnamefont {F.}~\bibnamefont
  {Vernizzi}},\ }\href {\doibase 10.1103/PhysRevLett.119.251302} {\bibfield
  {journal} {\bibinfo  {journal} {Phys. Rev. Lett.}\ }\textbf {\bibinfo
  {volume} {119}},\ \bibinfo {pages} {251302} (\bibinfo {year} {2017})},\
  \Eprint {http://arxiv.org/abs/1710.05877} {arXiv:1710.05877 [astro-ph.CO]}
  \BibitemShut {NoStop}%
%%CITATION = ARXIV:1710.05877;%%
\bibitem [{\citenamefont {Ezquiaga}\ and\ \citenamefont
  {Zumalacsrregui}(2017)}]{Ezquiaga:2017ekz}%
  \BibitemOpen
  \bibfield  {author} {\bibinfo {author} {\bibfnamefont {J.~M.}\ \bibnamefont
  {Ezquiaga}}\ and\ \bibinfo {author} {\bibfnamefont {M.}~\bibnamefont
  {Zumalacsrregui}},\ }\href {\doibase 10.1103/PhysRevLett.119.251304}
  {\bibfield  {journal} {\bibinfo  {journal} {Phys. Rev. Lett.}\ }\textbf
  {\bibinfo {volume} {119}},\ \bibinfo {pages} {251304} (\bibinfo {year}
  {2017})},\ \Eprint {http://arxiv.org/abs/1710.05901} {arXiv:1710.05901
  [astro-ph.CO]} \BibitemShut {NoStop}%
%%CITATION = ARXIV:1710.05901;%%
\bibitem [{\citenamefont {Baker}\ \emph {et~al.}(2017)\citenamefont {Baker},
  \citenamefont {Bellini}, \citenamefont {Ferreira}, \citenamefont {Lagos},
  \citenamefont {Noller},\ and\ \citenamefont {Sawicki}}]{Baker:2017hug}%
  \BibitemOpen
  \bibfield  {author} {\bibinfo {author} {\bibfnamefont {T.}~\bibnamefont
  {Baker}}, \bibinfo {author} {\bibfnamefont {E.}~\bibnamefont {Bellini}},
  \bibinfo {author} {\bibfnamefont {P.~G.}\ \bibnamefont {Ferreira}}, \bibinfo
  {author} {\bibfnamefont {M.}~\bibnamefont {Lagos}}, \bibinfo {author}
  {\bibfnamefont {J.}~\bibnamefont {Noller}}, \ and\ \bibinfo {author}
  {\bibfnamefont {I.}~\bibnamefont {Sawicki}},\ }\href {\doibase
  10.1103/PhysRevLett.119.251301} {\bibfield  {journal} {\bibinfo  {journal}
  {Phys. Rev. Lett.}\ }\textbf {\bibinfo {volume} {119}},\ \bibinfo {pages}
  {251301} (\bibinfo {year} {2017})},\ \Eprint
  {http://arxiv.org/abs/1710.06394} {arXiv:1710.06394 [astro-ph.CO]}
  \BibitemShut {NoStop}%
%%CITATION = ARXIV:1710.06394;%%
\bibitem [{\citenamefont {Sakstein}\ and\ \citenamefont
  {Jain}(2017)}]{Sakstein:2017xjx}%
  \BibitemOpen
  \bibfield  {author} {\bibinfo {author} {\bibfnamefont {J.}~\bibnamefont
  {Sakstein}}\ and\ \bibinfo {author} {\bibfnamefont {B.}~\bibnamefont
  {Jain}},\ }\href {\doibase 10.1103/PhysRevLett.119.251303} {\bibfield
  {journal} {\bibinfo  {journal} {Phys. Rev. Lett.}\ }\textbf {\bibinfo
  {volume} {119}},\ \bibinfo {pages} {251303} (\bibinfo {year} {2017})},\
  \Eprint {http://arxiv.org/abs/1710.05893} {arXiv:1710.05893 [astro-ph.CO]}
  \BibitemShut {NoStop}%
%%CITATION = ARXIV:1710.05893;%%
\bibitem [{\citenamefont {Bettoni}\ \emph {et~al.}(2017)\citenamefont
  {Bettoni}, \citenamefont {Ezquiaga}, \citenamefont {Hinterbichler},\ and\
  \citenamefont {Zumalacárregui}}]{Bettoni:2016mij}%
  \BibitemOpen
  \bibfield  {author} {\bibinfo {author} {\bibfnamefont {D.}~\bibnamefont
  {Bettoni}}, \bibinfo {author} {\bibfnamefont {J.~M.}\ \bibnamefont
  {Ezquiaga}}, \bibinfo {author} {\bibfnamefont {K.}~\bibnamefont
  {Hinterbichler}}, \ and\ \bibinfo {author} {\bibfnamefont {M.}~\bibnamefont
  {Zumalacárregui}},\ }\href {\doibase 10.1103/PhysRevD.95.084029} {\bibfield
  {journal} {\bibinfo  {journal} {Phys. Rev.}\ }\textbf {\bibinfo {volume}
  {D95}},\ \bibinfo {pages} {084029} (\bibinfo {year} {2017})},\ \Eprint
  {http://arxiv.org/abs/1608.01982} {arXiv:1608.01982 [gr-qc]} \BibitemShut
  {NoStop}%
%%CITATION = ARXIV:1608.01982;%%
\bibitem [{\citenamefont {Kase}\ and\ \citenamefont
  {Tsujikawa}(2019)}]{Kase:2018aps}%
  \BibitemOpen
  \bibfield  {author} {\bibinfo {author} {\bibfnamefont {R.}~\bibnamefont
  {Kase}}\ and\ \bibinfo {author} {\bibfnamefont {S.}~\bibnamefont
  {Tsujikawa}},\ }\href {\doibase 10.1142/S0218271819420057} {\bibfield
  {journal} {\bibinfo  {journal} {Int. J. Mod. Phys.}\ }\textbf {\bibinfo
  {volume} {D28}},\ \bibinfo {pages} {1942005} (\bibinfo {year} {2019})},\
  \Eprint {http://arxiv.org/abs/1809.08735} {arXiv:1809.08735 [gr-qc]}
  \BibitemShut {NoStop}%
%%CITATION = ARXIV:1809.08735;%%
\bibitem [{\citenamefont {Giacomello}\ \emph {et~al.}(2019)\citenamefont
  {Giacomello}, \citenamefont {De~Felice},\ and\ \citenamefont
  {Ansoldi}}]{Giacomello:2018jfi}%
  \BibitemOpen
  \bibfield  {author} {\bibinfo {author} {\bibfnamefont {F.}~\bibnamefont
  {Giacomello}}, \bibinfo {author} {\bibfnamefont {A.}~\bibnamefont
  {De~Felice}}, \ and\ \bibinfo {author} {\bibfnamefont {S.}~\bibnamefont
  {Ansoldi}},\ }\href {\doibase 10.1088/1475-7516/2019/03/038} {\bibfield
  {journal} {\bibinfo  {journal} {JCAP}\ }\textbf {\bibinfo {volume} {1903}},\
  \bibinfo {pages} {038} (\bibinfo {year} {2019})},\ \Eprint
  {http://arxiv.org/abs/1811.10885} {arXiv:1811.10885 [astro-ph.CO]}
  \BibitemShut {NoStop}%
%%CITATION = ARXIV:1811.10885;%%
\bibitem [{\citenamefont {Barreira}\ \emph {et~al.}(2014)\citenamefont
  {Barreira}, \citenamefont {Li}, \citenamefont {Baugh},\ and\ \citenamefont
  {Pascoli}}]{Barreira:2014jha}%
  \BibitemOpen
  \bibfield  {author} {\bibinfo {author} {\bibfnamefont {A.}~\bibnamefont
  {Barreira}}, \bibinfo {author} {\bibfnamefont {B.}~\bibnamefont {Li}},
  \bibinfo {author} {\bibfnamefont {C.}~\bibnamefont {Baugh}}, \ and\ \bibinfo
  {author} {\bibfnamefont {S.}~\bibnamefont {Pascoli}},\ }\href {\doibase
  10.1088/1475-7516/2014/08/059} {\bibfield  {journal} {\bibinfo  {journal}
  {JCAP}\ }\textbf {\bibinfo {volume} {1408}},\ \bibinfo {pages} {059}
  (\bibinfo {year} {2014})},\ \Eprint {http://arxiv.org/abs/1406.0485}
  {arXiv:1406.0485 [astro-ph.CO]} \BibitemShut {NoStop}%
%%CITATION = ARXIV:1406.0485;%%
\bibitem [{\citenamefont {Kobayashi}\ \emph
  {et~al.}(2010{\natexlab{a}})\citenamefont {Kobayashi}, \citenamefont
  {Tashiro},\ and\ \citenamefont {Suzuki}}]{Kobayashi:2009wr}%
  \BibitemOpen
  \bibfield  {author} {\bibinfo {author} {\bibfnamefont {T.}~\bibnamefont
  {Kobayashi}}, \bibinfo {author} {\bibfnamefont {H.}~\bibnamefont {Tashiro}},
  \ and\ \bibinfo {author} {\bibfnamefont {D.}~\bibnamefont {Suzuki}},\ }\href
  {\doibase 10.1103/PhysRevD.81.063513} {\bibfield  {journal} {\bibinfo
  {journal} {Phys. Rev.}\ }\textbf {\bibinfo {volume} {D81}},\ \bibinfo {pages}
  {063513} (\bibinfo {year} {2010}{\natexlab{a}})},\ \Eprint
  {http://arxiv.org/abs/0912.4641} {arXiv:0912.4641 [astro-ph.CO]} \BibitemShut
  {NoStop}%
%%CITATION = ARXIV:0912.4641;%%
\bibitem [{\citenamefont {Gubitosi}\ \emph {et~al.}(2013)\citenamefont
  {Gubitosi}, \citenamefont {Piazza},\ and\ \citenamefont
  {Vernizzi}}]{Gubitosi:2012hu}%
  \BibitemOpen
  \bibfield  {author} {\bibinfo {author} {\bibfnamefont {G.}~\bibnamefont
  {Gubitosi}}, \bibinfo {author} {\bibfnamefont {F.}~\bibnamefont {Piazza}}, \
  and\ \bibinfo {author} {\bibfnamefont {F.}~\bibnamefont {Vernizzi}},\ }\href
  {\doibase 10.1088/1475-7516/2013/02/032} {\bibfield  {journal} {\bibinfo
  {journal} {JCAP}\ }\textbf {\bibinfo {volume} {1302}},\ \bibinfo {pages}
  {032} (\bibinfo {year} {2013})},\ \bibinfo {note} {[JCAP1302,032(2013)]},\
  \Eprint {http://arxiv.org/abs/1210.0201} {arXiv:1210.0201 [hep-th]}
  \BibitemShut {NoStop}%
%%CITATION = ARXIV:1210.0201;%%
\bibitem [{\citenamefont {Bloomfield}\ \emph {et~al.}(2013)\citenamefont
  {Bloomfield}, \citenamefont {Flanagan}, \citenamefont {Park},\ and\
  \citenamefont {Watson}}]{Bloomfield:2012ff}%
  \BibitemOpen
  \bibfield  {author} {\bibinfo {author} {\bibfnamefont {J.~K.}\ \bibnamefont
  {Bloomfield}}, \bibinfo {author} {\bibfnamefont {E.~E.}\ \bibnamefont
  {Flanagan}}, \bibinfo {author} {\bibfnamefont {M.}~\bibnamefont {Park}}, \
  and\ \bibinfo {author} {\bibfnamefont {S.}~\bibnamefont {Watson}},\ }\href
  {\doibase 10.1088/1475-7516/2013/08/010} {\bibfield  {journal} {\bibinfo
  {journal} {JCAP}\ }\textbf {\bibinfo {volume} {1308}},\ \bibinfo {pages}
  {010} (\bibinfo {year} {2013})},\ \Eprint {http://arxiv.org/abs/1211.7054}
  {arXiv:1211.7054 [astro-ph.CO]} \BibitemShut {NoStop}%
%%CITATION = ARXIV:1211.7054;%%
\bibitem [{\citenamefont {Hu}\ \emph {et~al.}(2014{\natexlab{a}})\citenamefont
  {Hu}, \citenamefont {Raveri}, \citenamefont {Frusciante},\ and\ \citenamefont
  {Silvestri}}]{Hu:2013twa}%
  \BibitemOpen
  \bibfield  {author} {\bibinfo {author} {\bibfnamefont {B.}~\bibnamefont
  {Hu}}, \bibinfo {author} {\bibfnamefont {M.}~\bibnamefont {Raveri}}, \bibinfo
  {author} {\bibfnamefont {N.}~\bibnamefont {Frusciante}}, \ and\ \bibinfo
  {author} {\bibfnamefont {A.}~\bibnamefont {Silvestri}},\ }\href {\doibase
  10.1103/PhysRevD.89.103530} {\bibfield  {journal} {\bibinfo  {journal} {Phys.
  Rev.}\ }\textbf {\bibinfo {volume} {D89}},\ \bibinfo {pages} {103530}
  (\bibinfo {year} {2014}{\natexlab{a}})},\ \Eprint
  {http://arxiv.org/abs/1312.5742} {arXiv:1312.5742 [astro-ph.CO]} \BibitemShut
  {NoStop}%
%%CITATION = ARXIV:1312.5742;%%
\bibitem [{\citenamefont {Raveri}\ \emph {et~al.}(2014)\citenamefont {Raveri},
  \citenamefont {Hu}, \citenamefont {Frusciante},\ and\ \citenamefont
  {Silvestri}}]{Raveri:2014cka}%
  \BibitemOpen
  \bibfield  {author} {\bibinfo {author} {\bibfnamefont {M.}~\bibnamefont
  {Raveri}}, \bibinfo {author} {\bibfnamefont {B.}~\bibnamefont {Hu}}, \bibinfo
  {author} {\bibfnamefont {N.}~\bibnamefont {Frusciante}}, \ and\ \bibinfo
  {author} {\bibfnamefont {A.}~\bibnamefont {Silvestri}},\ }\href {\doibase
  10.1103/PhysRevD.90.043513} {\bibfield  {journal} {\bibinfo  {journal} {Phys.
  Rev.}\ }\textbf {\bibinfo {volume} {D90}},\ \bibinfo {pages} {043513}
  (\bibinfo {year} {2014})},\ \Eprint {http://arxiv.org/abs/1405.1022}
  {arXiv:1405.1022 [astro-ph.CO]} \BibitemShut {NoStop}%
%%CITATION = ARXIV:1405.1022;%%
\bibitem [{\citenamefont {Lesgourgues}\ and\ \citenamefont
  {Pastor}(2006)}]{Lesgourgues:2006nd}%
  \BibitemOpen
  \bibfield  {author} {\bibinfo {author} {\bibfnamefont {J.}~\bibnamefont
  {Lesgourgues}}\ and\ \bibinfo {author} {\bibfnamefont {S.}~\bibnamefont
  {Pastor}},\ }\href {\doibase 10.1016/j.physrep.2006.04.001} {\bibfield
  {journal} {\bibinfo  {journal} {Phys. Rept.}\ }\textbf {\bibinfo {volume}
  {429}},\ \bibinfo {pages} {307} (\bibinfo {year} {2006})},\ \Eprint
  {http://arxiv.org/abs/astro-ph/0603494} {arXiv:astro-ph/0603494 [astro-ph]}
  \BibitemShut {NoStop}%
%%CITATION = ASTRO-PH/0603494;%%
\bibitem [{\citenamefont {Wong}(2011)}]{Wong:2011ip}%
  \BibitemOpen
  \bibfield  {author} {\bibinfo {author} {\bibfnamefont {Y.~Y.~Y.}\
  \bibnamefont {Wong}},\ }\href {\doibase 10.1146/annurev-nucl-102010-130252}
  {\bibfield  {journal} {\bibinfo  {journal} {Ann. Rev. Nucl. Part. Sci.}\
  }\textbf {\bibinfo {volume} {61}},\ \bibinfo {pages} {69} (\bibinfo {year}
  {2011})},\ \Eprint {http://arxiv.org/abs/1111.1436} {arXiv:1111.1436
  [astro-ph.CO]} \BibitemShut {NoStop}%
%%CITATION = ARXIV:1111.1436;%%
\bibitem [{\citenamefont {Riess}\ \emph {et~al.}(2011)\citenamefont {Riess},
  \citenamefont {Macri}, \citenamefont {Casertano}, \citenamefont {Lampeitl},
  \citenamefont {Ferguson}, \citenamefont {Filippenko}, \citenamefont {Jha},
  \citenamefont {Li},\ and\ \citenamefont {Chornock}}]{Riess:2011yx}%
  \BibitemOpen
  \bibfield  {author} {\bibinfo {author} {\bibfnamefont {A.~G.}\ \bibnamefont
  {Riess}}, \bibinfo {author} {\bibfnamefont {L.}~\bibnamefont {Macri}},
  \bibinfo {author} {\bibfnamefont {S.}~\bibnamefont {Casertano}}, \bibinfo
  {author} {\bibfnamefont {H.}~\bibnamefont {Lampeitl}}, \bibinfo {author}
  {\bibfnamefont {H.~C.}\ \bibnamefont {Ferguson}}, \bibinfo {author}
  {\bibfnamefont {A.~V.}\ \bibnamefont {Filippenko}}, \bibinfo {author}
  {\bibfnamefont {S.~W.}\ \bibnamefont {Jha}}, \bibinfo {author} {\bibfnamefont
  {W.}~\bibnamefont {Li}}, \ and\ \bibinfo {author} {\bibfnamefont
  {R.}~\bibnamefont {Chornock}},\ }\href {\doibase 10.1088/0004-637X/732/2/129,
  10.1088/0004-637X/730/2/119} {\bibfield  {journal} {\bibinfo  {journal}
  {Astrophys. J.}\ }\textbf {\bibinfo {volume} {730}},\ \bibinfo {pages} {119}
  (\bibinfo {year} {2011})},\ \bibinfo {note} {[Erratum: Astrophys.
  J.732,129(2011)]},\ \Eprint {http://arxiv.org/abs/1103.2976} {arXiv:1103.2976
  [astro-ph.CO]} \BibitemShut {NoStop}%
%%CITATION = ARXIV:1103.2976;%%
\bibitem [{\citenamefont {Riess}\ \emph {et~al.}(2016)\citenamefont {Riess}
  \emph {et~al.}}]{Riess:2016jrr}%
  \BibitemOpen
  \bibfield  {author} {\bibinfo {author} {\bibfnamefont {A.~G.}\ \bibnamefont
  {Riess}} \emph {et~al.},\ }\href {\doibase 10.3847/0004-637X/826/1/56}
  {\bibfield  {journal} {\bibinfo  {journal} {Astrophys. J.}\ }\textbf
  {\bibinfo {volume} {826}},\ \bibinfo {pages} {56} (\bibinfo {year} {2016})},\
  \Eprint {http://arxiv.org/abs/1604.01424} {arXiv:1604.01424 [astro-ph.CO]}
  \BibitemShut {NoStop}%
%%CITATION = ARXIV:1604.01424;%%
\bibitem [{\citenamefont {Riess}\ \emph {et~al.}(2019)\citenamefont {Riess},
  \citenamefont {Casertano}, \citenamefont {Yuan}, \citenamefont {Macri},\ and\
  \citenamefont {Scolnic}}]{Riess:2019cxk}%
  \BibitemOpen
  \bibfield  {author} {\bibinfo {author} {\bibfnamefont {A.~G.}\ \bibnamefont
  {Riess}}, \bibinfo {author} {\bibfnamefont {S.}~\bibnamefont {Casertano}},
  \bibinfo {author} {\bibfnamefont {W.}~\bibnamefont {Yuan}}, \bibinfo {author}
  {\bibfnamefont {L.~M.}\ \bibnamefont {Macri}}, \ and\ \bibinfo {author}
  {\bibfnamefont {D.}~\bibnamefont {Scolnic}},\ }\href {\doibase
  10.3847/1538-4357/ab1422} {\bibfield  {journal} {\bibinfo  {journal}
  {Astrophys. J.}\ }\textbf {\bibinfo {volume} {876}},\ \bibinfo {pages} {85}
  (\bibinfo {year} {2019})},\ \Eprint {http://arxiv.org/abs/1903.07603}
  {arXiv:1903.07603 [astro-ph.CO]} \BibitemShut {NoStop}%
%%CITATION = ARXIV:1903.07603;%%
\bibitem [{\citenamefont {Delubac}\ \emph {et~al.}(2015)\citenamefont {Delubac}
  \emph {et~al.}}]{Delubac:2014aqe}%
  \BibitemOpen
  \bibfield  {author} {\bibinfo {author} {\bibfnamefont {T.}~\bibnamefont
  {Delubac}} \emph {et~al.} (\bibinfo {collaboration} {BOSS}),\ }\href
  {\doibase 10.1051/0004-6361/201423969} {\bibfield  {journal} {\bibinfo
  {journal} {Astron. Astrophys.}\ }\textbf {\bibinfo {volume} {574}},\ \bibinfo
  {pages} {A59} (\bibinfo {year} {2015})},\ \Eprint
  {http://arxiv.org/abs/1404.1801} {arXiv:1404.1801 [astro-ph.CO]} \BibitemShut
  {NoStop}%
%%CITATION = ARXIV:1404.1801;%%
\bibitem [{\citenamefont {Abbott}\ \emph
  {et~al.}(2018{\natexlab{a}})\citenamefont {Abbott} \emph
  {et~al.}}]{Abbott:2017wau}%
  \BibitemOpen
  \bibfield  {author} {\bibinfo {author} {\bibfnamefont {T.~M.~C.}\
  \bibnamefont {Abbott}} \emph {et~al.} (\bibinfo {collaboration} {DES}),\
  }\href {\doibase 10.1103/PhysRevD.98.043526} {\bibfield  {journal} {\bibinfo
  {journal} {Phys. Rev.}\ }\textbf {\bibinfo {volume} {D98}},\ \bibinfo {pages}
  {043526} (\bibinfo {year} {2018}{\natexlab{a}})},\ \Eprint
  {http://arxiv.org/abs/1708.01530} {arXiv:1708.01530 [astro-ph.CO]}
  \BibitemShut {NoStop}%
%%CITATION = ARXIV:1708.01530;%%
\bibitem [{\citenamefont {Abbott}\ \emph
  {et~al.}(2018{\natexlab{b}})\citenamefont {Abbott} \emph
  {et~al.}}]{Abbott:2017smn}%
  \BibitemOpen
  \bibfield  {author} {\bibinfo {author} {\bibfnamefont {T.~M.~C.}\
  \bibnamefont {Abbott}} \emph {et~al.} (\bibinfo {collaboration} {DES}),\
  }\href {\doibase 10.1093/mnras/sty1939} {\bibfield  {journal} {\bibinfo
  {journal} {Mon. Not. Roy. Astron. Soc.}\ }\textbf {\bibinfo {volume} {480}},\
  \bibinfo {pages} {3879} (\bibinfo {year} {2018}{\natexlab{b}})},\ \Eprint
  {http://arxiv.org/abs/1711.00403} {arXiv:1711.00403 [astro-ph.CO]}
  \BibitemShut {NoStop}%
%%CITATION = ARXIV:1711.00403;%%
\bibitem [{\citenamefont {Freedman}\ \emph {et~al.}(2019)\citenamefont
  {Freedman} \emph {et~al.}}]{Freedman:2019jwv}%
  \BibitemOpen
  \bibfield  {author} {\bibinfo {author} {\bibfnamefont {W.~L.}\ \bibnamefont
  {Freedman}} \emph {et~al.},\ }\href {\doibase 10.3847/1538-4357/ab2f73} {\
  (\bibinfo {year} {2019}),\ 10.3847/1538-4357/ab2f73},\ \Eprint
  {http://arxiv.org/abs/1907.05922} {arXiv:1907.05922 [astro-ph.CO]}
  \BibitemShut {NoStop}%
%%CITATION = ARXIV:1907.05922;%%
\bibitem [{\citenamefont {Yuan}\ \emph {et~al.}(2019)\citenamefont {Yuan},
  \citenamefont {Riess}, \citenamefont {Macri}, \citenamefont {Casertano},\
  and\ \citenamefont {Scolnic}}]{Yuan:2019npk}%
  \BibitemOpen
  \bibfield  {author} {\bibinfo {author} {\bibfnamefont {W.}~\bibnamefont
  {Yuan}}, \bibinfo {author} {\bibfnamefont {A.~G.}\ \bibnamefont {Riess}},
  \bibinfo {author} {\bibfnamefont {L.~M.}\ \bibnamefont {Macri}}, \bibinfo
  {author} {\bibfnamefont {S.}~\bibnamefont {Casertano}}, \ and\ \bibinfo
  {author} {\bibfnamefont {D.}~\bibnamefont {Scolnic}},\ }\href {\doibase
  10.3847/1538-4357/ab4bc9} {\bibfield  {journal} {\bibinfo  {journal}
  {Astrophys. J.}\ }\textbf {\bibinfo {volume} {886}},\ \bibinfo {pages} {61}
  (\bibinfo {year} {2019})},\ \Eprint {http://arxiv.org/abs/1908.00993}
  {arXiv:1908.00993 [astro-ph.GA]} \BibitemShut {NoStop}%
%%CITATION = ARXIV:1908.00993;%%
\bibitem [{\citenamefont {Karwal}\ and\ \citenamefont
  {Kamionkowski}(2016)}]{Karwal:2016vyq}%
  \BibitemOpen
  \bibfield  {author} {\bibinfo {author} {\bibfnamefont {T.}~\bibnamefont
  {Karwal}}\ and\ \bibinfo {author} {\bibfnamefont {M.}~\bibnamefont
  {Kamionkowski}},\ }\href {\doibase 10.1103/PhysRevD.94.103523} {\bibfield
  {journal} {\bibinfo  {journal} {Phys. Rev.}\ }\textbf {\bibinfo {volume}
  {D94}},\ \bibinfo {pages} {103523} (\bibinfo {year} {2016})},\ \Eprint
  {http://arxiv.org/abs/1608.01309} {arXiv:1608.01309 [astro-ph.CO]}
  \BibitemShut {NoStop}%
%%CITATION = ARXIV:1608.01309;%%
\bibitem [{\citenamefont {Pourtsidou}\ and\ \citenamefont
  {Tram}(2016)}]{Pourtsidou:2016ico}%
  \BibitemOpen
  \bibfield  {author} {\bibinfo {author} {\bibfnamefont {A.}~\bibnamefont
  {Pourtsidou}}\ and\ \bibinfo {author} {\bibfnamefont {T.}~\bibnamefont
  {Tram}},\ }\href {\doibase 10.1103/PhysRevD.94.043518} {\bibfield  {journal}
  {\bibinfo  {journal} {Phys. Rev.}\ }\textbf {\bibinfo {volume} {D94}},\
  \bibinfo {pages} {043518} (\bibinfo {year} {2016})},\ \Eprint
  {http://arxiv.org/abs/1604.04222} {arXiv:1604.04222 [astro-ph.CO]}
  \BibitemShut {NoStop}%
%%CITATION = ARXIV:1604.04222;%%
\bibitem [{\citenamefont {Poulin}\ \emph {et~al.}(2019)\citenamefont {Poulin},
  \citenamefont {Smith}, \citenamefont {Karwal},\ and\ \citenamefont
  {Kamionkowski}}]{Poulin:2018cxd}%
  \BibitemOpen
  \bibfield  {author} {\bibinfo {author} {\bibfnamefont {V.}~\bibnamefont
  {Poulin}}, \bibinfo {author} {\bibfnamefont {T.~L.}\ \bibnamefont {Smith}},
  \bibinfo {author} {\bibfnamefont {T.}~\bibnamefont {Karwal}}, \ and\ \bibinfo
  {author} {\bibfnamefont {M.}~\bibnamefont {Kamionkowski}},\ }\href {\doibase
  10.1103/PhysRevLett.122.221301} {\bibfield  {journal} {\bibinfo  {journal}
  {Phys. Rev. Lett.}\ }\textbf {\bibinfo {volume} {122}},\ \bibinfo {pages}
  {221301} (\bibinfo {year} {2019})},\ \Eprint
  {http://arxiv.org/abs/1811.04083} {arXiv:1811.04083 [astro-ph.CO]}
  \BibitemShut {NoStop}%
%%CITATION = ARXIV:1811.04083;%%
\bibitem [{\citenamefont {Lin}\ \emph {et~al.}(2019)\citenamefont {Lin},
  \citenamefont {Raveri},\ and\ \citenamefont {Hu}}]{Lin:2018nxe}%
  \BibitemOpen
  \bibfield  {author} {\bibinfo {author} {\bibfnamefont {M.-X.}\ \bibnamefont
  {Lin}}, \bibinfo {author} {\bibfnamefont {M.}~\bibnamefont {Raveri}}, \ and\
  \bibinfo {author} {\bibfnamefont {W.}~\bibnamefont {Hu}},\ }\href {\doibase
  10.1103/PhysRevD.99.043514} {\bibfield  {journal} {\bibinfo  {journal} {Phys.
  Rev.}\ }\textbf {\bibinfo {volume} {D99}},\ \bibinfo {pages} {043514}
  (\bibinfo {year} {2019})},\ \Eprint {http://arxiv.org/abs/1810.02333}
  {arXiv:1810.02333 [astro-ph.CO]} \BibitemShut {NoStop}%
%%CITATION = ARXIV:1810.02333;%%
\bibitem [{\citenamefont {Agrawal}\ \emph
  {et~al.}(2019{\natexlab{a}})\citenamefont {Agrawal}, \citenamefont {Obied},\
  and\ \citenamefont {Vafa}}]{Agrawal:2019dlm}%
  \BibitemOpen
  \bibfield  {author} {\bibinfo {author} {\bibfnamefont {P.}~\bibnamefont
  {Agrawal}}, \bibinfo {author} {\bibfnamefont {G.}~\bibnamefont {Obied}}, \
  and\ \bibinfo {author} {\bibfnamefont {C.}~\bibnamefont {Vafa}},\ }\href@noop
  {} {\  (\bibinfo {year} {2019}{\natexlab{a}})},\ \Eprint
  {http://arxiv.org/abs/1906.08261} {arXiv:1906.08261 [astro-ph.CO]}
  \BibitemShut {NoStop}%
%%CITATION = ARXIV:1906.08261;%%
\bibitem [{\citenamefont {Agrawal}\ \emph
  {et~al.}(2019{\natexlab{b}})\citenamefont {Agrawal}, \citenamefont
  {Cyr-Racine}, \citenamefont {Pinner},\ and\ \citenamefont
  {Randall}}]{Agrawal:2019lmo}%
  \BibitemOpen
  \bibfield  {author} {\bibinfo {author} {\bibfnamefont {P.}~\bibnamefont
  {Agrawal}}, \bibinfo {author} {\bibfnamefont {F.-Y.}\ \bibnamefont
  {Cyr-Racine}}, \bibinfo {author} {\bibfnamefont {D.}~\bibnamefont {Pinner}},
  \ and\ \bibinfo {author} {\bibfnamefont {L.}~\bibnamefont {Randall}},\
  }\href@noop {} {\  (\bibinfo {year} {2019}{\natexlab{b}})},\ \Eprint
  {http://arxiv.org/abs/1904.01016} {arXiv:1904.01016 [astro-ph.CO]}
  \BibitemShut {NoStop}%
%%CITATION = ARXIV:1904.01016;%%
\bibitem [{\citenamefont {Kaloper}(2019)}]{Kaloper:2019lpl}%
  \BibitemOpen
  \bibfield  {author} {\bibinfo {author} {\bibfnamefont {N.}~\bibnamefont
  {Kaloper}},\ }\href {\doibase 10.1142/S0218271819440176} {\bibfield
  {journal} {\bibinfo  {journal} {Int. J. Phys.}\ }\textbf {\bibinfo {volume}
  {D}} (\bibinfo {year} {2019}),\ 10.1142/S0218271819440176},\ \Eprint
  {http://arxiv.org/abs/1903.11676} {arXiv:1903.11676 [hep-th]} \BibitemShut
  {NoStop}%
%%CITATION = ARXIV:1903.11676;%%
\bibitem [{\citenamefont {Desmond}\ \emph {et~al.}(2019)\citenamefont
  {Desmond}, \citenamefont {Jain},\ and\ \citenamefont
  {Sakstein}}]{Desmond:2019ygn}%
  \BibitemOpen
  \bibfield  {author} {\bibinfo {author} {\bibfnamefont {H.}~\bibnamefont
  {Desmond}}, \bibinfo {author} {\bibfnamefont {B.}~\bibnamefont {Jain}}, \
  and\ \bibinfo {author} {\bibfnamefont {J.}~\bibnamefont {Sakstein}},\ }\href
  {\doibase 10.1103/PhysRevD.100.043537} {\bibfield  {journal} {\bibinfo
  {journal} {Phys. Rev.}\ }\textbf {\bibinfo {volume} {D100}},\ \bibinfo
  {pages} {043537} (\bibinfo {year} {2019})},\ \Eprint
  {http://arxiv.org/abs/1907.03778} {arXiv:1907.03778 [astro-ph.CO]}
  \BibitemShut {NoStop}%
%%CITATION = ARXIV:1907.03778;%%
\bibitem [{\citenamefont {Di~Valentino}\ \emph {et~al.}(2019)\citenamefont
  {Di~Valentino}, \citenamefont {Melchiorri}, \citenamefont {Mena},\ and\
  \citenamefont {Vagnozzi}}]{DiValentino:2019jae}%
  \BibitemOpen
  \bibfield  {author} {\bibinfo {author} {\bibfnamefont {E.}~\bibnamefont
  {Di~Valentino}}, \bibinfo {author} {\bibfnamefont {A.}~\bibnamefont
  {Melchiorri}}, \bibinfo {author} {\bibfnamefont {O.}~\bibnamefont {Mena}}, \
  and\ \bibinfo {author} {\bibfnamefont {S.}~\bibnamefont {Vagnozzi}},\
  }\href@noop {} {\  (\bibinfo {year} {2019})},\ \Eprint
  {http://arxiv.org/abs/1910.09853} {arXiv:1910.09853 [astro-ph.CO]}
  \BibitemShut {NoStop}%
%%CITATION = ARXIV:1910.09853;%%
\bibitem [{\citenamefont {Sakstein}\ and\ \citenamefont
  {Trodden}(2019)}]{Sakstein:2019fmf}%
  \BibitemOpen
  \bibfield  {author} {\bibinfo {author} {\bibfnamefont {J.}~\bibnamefont
  {Sakstein}}\ and\ \bibinfo {author} {\bibfnamefont {M.}~\bibnamefont
  {Trodden}},\ }\href@noop {} {\  (\bibinfo {year} {2019})},\ \Eprint
  {http://arxiv.org/abs/1911.11760} {arXiv:1911.11760 [astro-ph.CO]}
  \BibitemShut {NoStop}%
%%CITATION = ARXIV:1911.11760;%%
\bibitem [{\citenamefont {Lombriser}(2019)}]{Lombriser:2019ahl}%
  \BibitemOpen
  \bibfield  {author} {\bibinfo {author} {\bibfnamefont {L.}~\bibnamefont
  {Lombriser}},\ }\href@noop {} {\  (\bibinfo {year} {2019})},\ \Eprint
  {http://arxiv.org/abs/1906.12347} {arXiv:1906.12347 [astro-ph.CO]}
  \BibitemShut {NoStop}%
%%CITATION = ARXIV:1906.12347;%%
\bibitem [{\citenamefont {Deffayet}\ \emph {et~al.}(2010)\citenamefont
  {Deffayet}, \citenamefont {Pujolas}, \citenamefont {Sawicki},\ and\
  \citenamefont {Vikman}}]{Deffayet:2010qz}%
  \BibitemOpen
  \bibfield  {author} {\bibinfo {author} {\bibfnamefont {C.}~\bibnamefont
  {Deffayet}}, \bibinfo {author} {\bibfnamefont {O.}~\bibnamefont {Pujolas}},
  \bibinfo {author} {\bibfnamefont {I.}~\bibnamefont {Sawicki}}, \ and\
  \bibinfo {author} {\bibfnamefont {A.}~\bibnamefont {Vikman}},\ }\href
  {\doibase 10.1088/1475-7516/2010/10/026} {\bibfield  {journal} {\bibinfo
  {journal} {JCAP}\ }\textbf {\bibinfo {volume} {1010}},\ \bibinfo {pages}
  {026} (\bibinfo {year} {2010})},\ \Eprint {http://arxiv.org/abs/1008.0048}
  {arXiv:1008.0048 [hep-th]} \BibitemShut {NoStop}%
%%CITATION = ARXIV:1008.0048;%%
\bibitem [{\citenamefont {Kobayashi}\ \emph
  {et~al.}(2010{\natexlab{b}})\citenamefont {Kobayashi}, \citenamefont
  {Yamaguchi},\ and\ \citenamefont {Yokoyama}}]{Kobayashi:2010cm}%
  \BibitemOpen
  \bibfield  {author} {\bibinfo {author} {\bibfnamefont {T.}~\bibnamefont
  {Kobayashi}}, \bibinfo {author} {\bibfnamefont {M.}~\bibnamefont
  {Yamaguchi}}, \ and\ \bibinfo {author} {\bibfnamefont {J.}~\bibnamefont
  {Yokoyama}},\ }\href {\doibase 10.1103/PhysRevLett.105.231302} {\bibfield
  {journal} {\bibinfo  {journal} {Phys. Rev. Lett.}\ }\textbf {\bibinfo
  {volume} {105}},\ \bibinfo {pages} {231302} (\bibinfo {year}
  {2010}{\natexlab{b}})},\ \Eprint {http://arxiv.org/abs/1008.0603}
  {arXiv:1008.0603 [hep-th]} \BibitemShut {NoStop}%
%%CITATION = ARXIV:1008.0603;%%
\bibitem [{\citenamefont {Barreira}\ \emph {et~al.}(2013)\citenamefont
  {Barreira}, \citenamefont {Li}, \citenamefont {Baugh},\ and\ \citenamefont
  {Pascoli}}]{Barreira:2013xea}%
  \BibitemOpen
  \bibfield  {author} {\bibinfo {author} {\bibfnamefont {A.}~\bibnamefont
  {Barreira}}, \bibinfo {author} {\bibfnamefont {B.}~\bibnamefont {Li}},
  \bibinfo {author} {\bibfnamefont {C.~M.}\ \bibnamefont {Baugh}}, \ and\
  \bibinfo {author} {\bibfnamefont {S.}~\bibnamefont {Pascoli}},\ }\href
  {\doibase 10.1088/1475-7516/2013/11/056} {\bibfield  {journal} {\bibinfo
  {journal} {JCAP}\ }\textbf {\bibinfo {volume} {1311}},\ \bibinfo {pages}
  {056} (\bibinfo {year} {2013})},\ \Eprint {http://arxiv.org/abs/1308.3699}
  {arXiv:1308.3699 [astro-ph.CO]} \BibitemShut {NoStop}%
%%CITATION = ARXIV:1308.3699;%%
\bibitem [{\citenamefont {{Barreira}}\ \emph {et~al.}(2014)\citenamefont
  {{Barreira}}, \citenamefont {{Li}}, \citenamefont {{Baugh}},\ and\
  \citenamefont {{Pascoli}}}]{2014JCAP...08..059B}%
  \BibitemOpen
  \bibfield  {author} {\bibinfo {author} {\bibfnamefont {A.}~\bibnamefont
  {{Barreira}}}, \bibinfo {author} {\bibfnamefont {B.}~\bibnamefont {{Li}}},
  \bibinfo {author} {\bibfnamefont {C.~M.}\ \bibnamefont {{Baugh}}}, \ and\
  \bibinfo {author} {\bibfnamefont {S.}~\bibnamefont {{Pascoli}}},\ }\href
  {\doibase 10.1088/1475-7516/2014/08/059} {\bibfield  {journal} {\bibinfo
  {journal} {jcap}\ }\textbf {\bibinfo {volume} {8}},\ \bibinfo {eid} {059}
  (\bibinfo {year} {2014})},\ \Eprint {http://arxiv.org/abs/1406.0485}
  {arXiv:1406.0485} \BibitemShut {NoStop}%
\bibitem [{\citenamefont {Gleyzes}\ \emph {et~al.}(2013)\citenamefont
  {Gleyzes}, \citenamefont {Langlois}, \citenamefont {Piazza},\ and\
  \citenamefont {Vernizzi}}]{Gleyzes:2013ooa}%
  \BibitemOpen
  \bibfield  {author} {\bibinfo {author} {\bibfnamefont {J.}~\bibnamefont
  {Gleyzes}}, \bibinfo {author} {\bibfnamefont {D.}~\bibnamefont {Langlois}},
  \bibinfo {author} {\bibfnamefont {F.}~\bibnamefont {Piazza}}, \ and\ \bibinfo
  {author} {\bibfnamefont {F.}~\bibnamefont {Vernizzi}},\ }\href {\doibase
  10.1088/1475-7516/2013/08/025} {\bibfield  {journal} {\bibinfo  {journal}
  {JCAP}\ }\textbf {\bibinfo {volume} {1308}},\ \bibinfo {pages} {025}
  (\bibinfo {year} {2013})},\ \Eprint {http://arxiv.org/abs/1304.4840}
  {arXiv:1304.4840 [hep-th]} \BibitemShut {NoStop}%
%%CITATION = ARXIV:1304.4840;%%
\bibitem [{\citenamefont {Piazza}\ \emph {et~al.}(2014)\citenamefont {Piazza},
  \citenamefont {Steigerwald},\ and\ \citenamefont
  {Marinoni}}]{Piazza:2013pua}%
  \BibitemOpen
  \bibfield  {author} {\bibinfo {author} {\bibfnamefont {F.}~\bibnamefont
  {Piazza}}, \bibinfo {author} {\bibfnamefont {H.}~\bibnamefont {Steigerwald}},
  \ and\ \bibinfo {author} {\bibfnamefont {C.}~\bibnamefont {Marinoni}},\
  }\href {\doibase 10.1088/1475-7516/2014/05/043} {\bibfield  {journal}
  {\bibinfo  {journal} {JCAP}\ }\textbf {\bibinfo {volume} {1405}},\ \bibinfo
  {pages} {043} (\bibinfo {year} {2014})},\ \Eprint
  {http://arxiv.org/abs/1312.6111} {arXiv:1312.6111 [astro-ph.CO]} \BibitemShut
  {NoStop}%
%%CITATION = ARXIV:1312.6111;%%
\bibitem [{\citenamefont {Tsujikawa}(2015)}]{Tsujikawa:2014mba}%
  \BibitemOpen
  \bibfield  {author} {\bibinfo {author} {\bibfnamefont {S.}~\bibnamefont
  {Tsujikawa}},\ }\href {\doibase 10.1007/978-3-319-10070-8_4} {\bibfield
  {journal} {\bibinfo  {journal} {Lect. Notes Phys.}\ }\textbf {\bibinfo
  {volume} {892}},\ \bibinfo {pages} {97} (\bibinfo {year} {2015})},\ \Eprint
  {http://arxiv.org/abs/1404.2684} {arXiv:1404.2684 [gr-qc]} \BibitemShut
  {NoStop}%
%%CITATION = ARXIV:1404.2684;%%
\bibitem [{\citenamefont {Frusciante}\ and\ \citenamefont
  {Perenon}(2019)}]{Frusciante:2019xia}%
  \BibitemOpen
  \bibfield  {author} {\bibinfo {author} {\bibfnamefont {N.}~\bibnamefont
  {Frusciante}}\ and\ \bibinfo {author} {\bibfnamefont {L.}~\bibnamefont
  {Perenon}},\ }\href@noop {} {\  (\bibinfo {year} {2019})},\ \Eprint
  {http://arxiv.org/abs/1907.03150} {arXiv:1907.03150 [astro-ph.CO]}
  \BibitemShut {NoStop}%
%%CITATION = ARXIV:1907.03150;%%
\bibitem [{\citenamefont {Hu}\ \emph {et~al.}(2014{\natexlab{b}})\citenamefont
  {Hu}, \citenamefont {Raveri}, \citenamefont {Frusciante},\ and\ \citenamefont
  {Silvestri}}]{Hu:2014oga}%
  \BibitemOpen
  \bibfield  {author} {\bibinfo {author} {\bibfnamefont {B.}~\bibnamefont
  {Hu}}, \bibinfo {author} {\bibfnamefont {M.}~\bibnamefont {Raveri}}, \bibinfo
  {author} {\bibfnamefont {N.}~\bibnamefont {Frusciante}}, \ and\ \bibinfo
  {author} {\bibfnamefont {A.}~\bibnamefont {Silvestri}},\ }\href@noop {} {\
  (\bibinfo {year} {2014}{\natexlab{b}})},\ \Eprint
  {http://arxiv.org/abs/1405.3590} {arXiv:1405.3590 [astro-ph.IM]} \BibitemShut
  {NoStop}%
%%CITATION = ARXIV:1405.3590;%%
\bibitem [{\citenamefont {Bloomfield}(2013)}]{Bloomfield:2013efa}%
  \BibitemOpen
  \bibfield  {author} {\bibinfo {author} {\bibfnamefont {J.}~\bibnamefont
  {Bloomfield}},\ }\href {\doibase 10.1088/1475-7516/2013/12/044} {\bibfield
  {journal} {\bibinfo  {journal} {JCAP}\ }\textbf {\bibinfo {volume} {1312}},\
  \bibinfo {pages} {044} (\bibinfo {year} {2013})},\ \Eprint
  {http://arxiv.org/abs/1304.6712} {arXiv:1304.6712 [astro-ph.CO]} \BibitemShut
  {NoStop}%
%%CITATION = ARXIV:1304.6712;%%
\bibitem [{\citenamefont {Gleyzes}\ \emph
  {et~al.}(2015{\natexlab{a}})\citenamefont {Gleyzes}, \citenamefont
  {Langlois},\ and\ \citenamefont {Vernizzi}}]{Gleyzes:2014rba}%
  \BibitemOpen
  \bibfield  {author} {\bibinfo {author} {\bibfnamefont {J.}~\bibnamefont
  {Gleyzes}}, \bibinfo {author} {\bibfnamefont {D.}~\bibnamefont {Langlois}}, \
  and\ \bibinfo {author} {\bibfnamefont {F.}~\bibnamefont {Vernizzi}},\ }\href
  {\doibase 10.1142/S021827181443010X} {\bibfield  {journal} {\bibinfo
  {journal} {Int. J. Mod. Phys.}\ }\textbf {\bibinfo {volume} {D23}},\ \bibinfo
  {pages} {1443010} (\bibinfo {year} {2015}{\natexlab{a}})},\ \Eprint
  {http://arxiv.org/abs/1411.3712} {arXiv:1411.3712 [hep-th]} \BibitemShut
  {NoStop}%
%%CITATION = ARXIV:1411.3712;%%
\bibitem [{\citenamefont {Frusciante}\ \emph
  {et~al.}(2016{\natexlab{a}})\citenamefont {Frusciante}, \citenamefont
  {Raveri}, \citenamefont {Vernieri}, \citenamefont {Hu},\ and\ \citenamefont
  {Silvestri}}]{Frusciante:2015maa}%
  \BibitemOpen
  \bibfield  {author} {\bibinfo {author} {\bibfnamefont {N.}~\bibnamefont
  {Frusciante}}, \bibinfo {author} {\bibfnamefont {M.}~\bibnamefont {Raveri}},
  \bibinfo {author} {\bibfnamefont {D.}~\bibnamefont {Vernieri}}, \bibinfo
  {author} {\bibfnamefont {B.}~\bibnamefont {Hu}}, \ and\ \bibinfo {author}
  {\bibfnamefont {A.}~\bibnamefont {Silvestri}},\ }\href {\doibase
  10.1016/j.dark.2016.03.002} {\bibfield  {journal} {\bibinfo  {journal} {Phys.
  Dark Univ.}\ }\textbf {\bibinfo {volume} {13}},\ \bibinfo {pages} {7}
  (\bibinfo {year} {2016}{\natexlab{a}})},\ \Eprint
  {http://arxiv.org/abs/1508.01787} {arXiv:1508.01787 [astro-ph.CO]}
  \BibitemShut {NoStop}%
%%CITATION = ARXIV:1508.01787;%%
\bibitem [{\citenamefont {Frusciante}\ \emph
  {et~al.}(2016{\natexlab{b}})\citenamefont {Frusciante}, \citenamefont
  {Papadomanolakis},\ and\ \citenamefont {Silvestri}}]{Frusciante:2016xoj}%
  \BibitemOpen
  \bibfield  {author} {\bibinfo {author} {\bibfnamefont {N.}~\bibnamefont
  {Frusciante}}, \bibinfo {author} {\bibfnamefont {G.}~\bibnamefont
  {Papadomanolakis}}, \ and\ \bibinfo {author} {\bibfnamefont {A.}~\bibnamefont
  {Silvestri}},\ }\href {\doibase 10.1088/1475-7516/2016/07/018} {\bibfield
  {journal} {\bibinfo  {journal} {JCAP}\ }\textbf {\bibinfo {volume} {1607}},\
  \bibinfo {pages} {018} (\bibinfo {year} {2016}{\natexlab{b}})},\ \Eprint
  {http://arxiv.org/abs/1601.04064} {arXiv:1601.04064 [gr-qc]} \BibitemShut
  {NoStop}%
%%CITATION = ARXIV:1601.04064;%%
\bibitem [{\citenamefont {Frusciante}\ \emph
  {et~al.}(2019{\natexlab{b}})\citenamefont {Frusciante}, \citenamefont
  {Peirone}, \citenamefont {Casas},\ and\ \citenamefont
  {Lima}}]{Frusciante:2018jzw}%
  \BibitemOpen
  \bibfield  {author} {\bibinfo {author} {\bibfnamefont {N.}~\bibnamefont
  {Frusciante}}, \bibinfo {author} {\bibfnamefont {S.}~\bibnamefont {Peirone}},
  \bibinfo {author} {\bibfnamefont {S.}~\bibnamefont {Casas}}, \ and\ \bibinfo
  {author} {\bibfnamefont {N.~A.}\ \bibnamefont {Lima}},\ }\href {\doibase
  10.1103/PhysRevD.99.063538} {\bibfield  {journal} {\bibinfo  {journal} {Phys.
  Rev.}\ }\textbf {\bibinfo {volume} {D99}},\ \bibinfo {pages} {063538}
  (\bibinfo {year} {2019}{\natexlab{b}})},\ \Eprint
  {http://arxiv.org/abs/1810.10521} {arXiv:1810.10521 [astro-ph.CO]}
  \BibitemShut {NoStop}%
%%CITATION = ARXIV:1810.10521;%%
\bibitem [{\citenamefont {Gleyzes}\ \emph
  {et~al.}(2015{\natexlab{b}})\citenamefont {Gleyzes}, \citenamefont
  {Langlois}, \citenamefont {Piazza},\ and\ \citenamefont
  {Vernizzi}}]{Gleyzes:2014qga}%
  \BibitemOpen
  \bibfield  {author} {\bibinfo {author} {\bibfnamefont {J.}~\bibnamefont
  {Gleyzes}}, \bibinfo {author} {\bibfnamefont {D.}~\bibnamefont {Langlois}},
  \bibinfo {author} {\bibfnamefont {F.}~\bibnamefont {Piazza}}, \ and\ \bibinfo
  {author} {\bibfnamefont {F.}~\bibnamefont {Vernizzi}},\ }\href {\doibase
  10.1088/1475-7516/2015/02/018} {\bibfield  {journal} {\bibinfo  {journal}
  {JCAP}\ }\textbf {\bibinfo {volume} {1502}},\ \bibinfo {pages} {018}
  (\bibinfo {year} {2015}{\natexlab{b}})},\ \Eprint
  {http://arxiv.org/abs/1408.1952} {arXiv:1408.1952 [astro-ph.CO]} \BibitemShut
  {NoStop}%
%%CITATION = ARXIV:1408.1952;%%
\bibitem [{\citenamefont {Gergely}\ and\ \citenamefont
  {Tsujikawa}(2014)}]{Gergely:2014rna}%
  \BibitemOpen
  \bibfield  {author} {\bibinfo {author} {\bibfnamefont {L.~A.}\ \bibnamefont
  {Gergely}}\ and\ \bibinfo {author} {\bibfnamefont {S.}~\bibnamefont
  {Tsujikawa}},\ }\href {\doibase 10.1103/PhysRevD.89.064059} {\bibfield
  {journal} {\bibinfo  {journal} {Phys. Rev.}\ }\textbf {\bibinfo {volume}
  {D89}},\ \bibinfo {pages} {064059} (\bibinfo {year} {2014})},\ \Eprint
  {http://arxiv.org/abs/1402.0553} {arXiv:1402.0553 [hep-th]} \BibitemShut
  {NoStop}%
%%CITATION = ARXIV:1402.0553;%%
\bibitem [{\citenamefont {Kase}\ and\ \citenamefont
  {Tsujikawa}(2014)}]{Kase:2014yya}%
  \BibitemOpen
  \bibfield  {author} {\bibinfo {author} {\bibfnamefont {R.}~\bibnamefont
  {Kase}}\ and\ \bibinfo {author} {\bibfnamefont {S.}~\bibnamefont
  {Tsujikawa}},\ }\href {\doibase 10.1103/PhysRevD.90.044073} {\bibfield
  {journal} {\bibinfo  {journal} {Phys. Rev.}\ }\textbf {\bibinfo {volume}
  {D90}},\ \bibinfo {pages} {044073} (\bibinfo {year} {2014})},\ \Eprint
  {http://arxiv.org/abs/1407.0794} {arXiv:1407.0794 [hep-th]} \BibitemShut
  {NoStop}%
%%CITATION = ARXIV:1407.0794;%%
\bibitem [{\citenamefont {De~Felice}\ \emph {et~al.}(2015)\citenamefont
  {De~Felice}, \citenamefont {Koyama},\ and\ \citenamefont
  {Tsujikawa}}]{DeFelice:2015isa}%
  \BibitemOpen
  \bibfield  {author} {\bibinfo {author} {\bibfnamefont {A.}~\bibnamefont
  {De~Felice}}, \bibinfo {author} {\bibfnamefont {K.}~\bibnamefont {Koyama}}, \
  and\ \bibinfo {author} {\bibfnamefont {S.}~\bibnamefont {Tsujikawa}},\ }\href
  {\doibase 10.1088/1475-7516/2015/05/058} {\bibfield  {journal} {\bibinfo
  {journal} {JCAP}\ }\textbf {\bibinfo {volume} {1505}},\ \bibinfo {pages}
  {058} (\bibinfo {year} {2015})},\ \Eprint {http://arxiv.org/abs/1503.06539}
  {arXiv:1503.06539 [gr-qc]} \BibitemShut {NoStop}%
%%CITATION = ARXIV:1503.06539;%%
\bibitem [{\citenamefont {De~Felice}\ \emph {et~al.}(2017)\citenamefont
  {De~Felice}, \citenamefont {Frusciante},\ and\ \citenamefont
  {Papadomanolakis}}]{DeFelice:2016ucp}%
  \BibitemOpen
  \bibfield  {author} {\bibinfo {author} {\bibfnamefont {A.}~\bibnamefont
  {De~Felice}}, \bibinfo {author} {\bibfnamefont {N.}~\bibnamefont
  {Frusciante}}, \ and\ \bibinfo {author} {\bibfnamefont {G.}~\bibnamefont
  {Papadomanolakis}},\ }\href {\doibase 10.1088/1475-7516/2017/03/027}
  {\bibfield  {journal} {\bibinfo  {journal} {JCAP}\ }\textbf {\bibinfo
  {volume} {1703}},\ \bibinfo {pages} {027} (\bibinfo {year} {2017})},\ \Eprint
  {http://arxiv.org/abs/1609.03599} {arXiv:1609.03599 [gr-qc]} \BibitemShut
  {NoStop}%
%%CITATION = ARXIV:1609.03599;%%
\bibitem [{\citenamefont {Frusciante}\ \emph
  {et~al.}(2019{\natexlab{c}})\citenamefont {Frusciante}, \citenamefont
  {Papadomanolakis}, \citenamefont {Peirone},\ and\ \citenamefont
  {Silvestri}}]{Frusciante:2018vht}%
  \BibitemOpen
  \bibfield  {author} {\bibinfo {author} {\bibfnamefont {N.}~\bibnamefont
  {Frusciante}}, \bibinfo {author} {\bibfnamefont {G.}~\bibnamefont
  {Papadomanolakis}}, \bibinfo {author} {\bibfnamefont {S.}~\bibnamefont
  {Peirone}}, \ and\ \bibinfo {author} {\bibfnamefont {A.}~\bibnamefont
  {Silvestri}},\ }\href {\doibase 10.1088/1475-7516/2019/02/029} {\bibfield
  {journal} {\bibinfo  {journal} {JCAP}\ }\textbf {\bibinfo {volume} {1902}},\
  \bibinfo {pages} {029} (\bibinfo {year} {2019}{\natexlab{c}})},\ \Eprint
  {http://arxiv.org/abs/1810.03461} {arXiv:1810.03461 [gr-qc]} \BibitemShut
  {NoStop}%
%%CITATION = ARXIV:1810.03461;%%
\bibitem [{\citenamefont {Bean}\ and\ \citenamefont
  {Tangmatitham}(2010)}]{Bean:2010zq}%
  \BibitemOpen
  \bibfield  {author} {\bibinfo {author} {\bibfnamefont {R.}~\bibnamefont
  {Bean}}\ and\ \bibinfo {author} {\bibfnamefont {M.}~\bibnamefont
  {Tangmatitham}},\ }\href {\doibase 10.1103/PhysRevD.81.083534} {\bibfield
  {journal} {\bibinfo  {journal} {Phys. Rev.}\ }\textbf {\bibinfo {volume}
  {D81}},\ \bibinfo {pages} {083534} (\bibinfo {year} {2010})},\ \Eprint
  {http://arxiv.org/abs/1002.4197} {arXiv:1002.4197 [astro-ph.CO]} \BibitemShut
  {NoStop}%
%%CITATION = ARXIV:1002.4197;%%
\bibitem [{\citenamefont {Silvestri}\ \emph {et~al.}(2013)\citenamefont
  {Silvestri}, \citenamefont {Pogosian},\ and\ \citenamefont
  {Buniy}}]{Silvestri:2013ne}%
  \BibitemOpen
  \bibfield  {author} {\bibinfo {author} {\bibfnamefont {A.}~\bibnamefont
  {Silvestri}}, \bibinfo {author} {\bibfnamefont {L.}~\bibnamefont {Pogosian}},
  \ and\ \bibinfo {author} {\bibfnamefont {R.~V.}\ \bibnamefont {Buniy}},\
  }\href {\doibase 10.1103/PhysRevD.87.104015} {\bibfield  {journal} {\bibinfo
  {journal} {Phys. Rev.}\ }\textbf {\bibinfo {volume} {D87}},\ \bibinfo {pages}
  {104015} (\bibinfo {year} {2013})},\ \Eprint {http://arxiv.org/abs/1302.1193}
  {arXiv:1302.1193 [astro-ph.CO]} \BibitemShut {NoStop}%
%%CITATION = ARXIV:1302.1193;%%
\bibitem [{\citenamefont {Aghanim}\ \emph {et~al.}(2016)\citenamefont {Aghanim}
  \emph {et~al.}}]{Aghanim:2015xee}%
  \BibitemOpen
  \bibfield  {author} {\bibinfo {author} {\bibfnamefont {N.}~\bibnamefont
  {Aghanim}} \emph {et~al.} (\bibinfo {collaboration} {Planck}),\ }\href
  {\doibase 10.1051/0004-6361/201526926} {\bibfield  {journal} {\bibinfo
  {journal} {Astron. Astrophys.}\ }\textbf {\bibinfo {volume} {594}},\ \bibinfo
  {pages} {A11} (\bibinfo {year} {2016})},\ \Eprint
  {http://arxiv.org/abs/1507.02704} {arXiv:1507.02704 [astro-ph.CO]}
  \BibitemShut {NoStop}%
%%CITATION = ARXIV:1507.02704;%%
\bibitem [{\citenamefont {Ade}\ \emph {et~al.}(2016)\citenamefont {Ade} \emph
  {et~al.}}]{Ade:2015xua}%
  \BibitemOpen
  \bibfield  {author} {\bibinfo {author} {\bibfnamefont {P.~A.~R.}\
  \bibnamefont {Ade}} \emph {et~al.} (\bibinfo {collaboration} {Planck}),\
  }\href {\doibase 10.1051/0004-6361/201525830} {\bibfield  {journal} {\bibinfo
   {journal} {Astron. Astrophys.}\ }\textbf {\bibinfo {volume} {594}},\
  \bibinfo {pages} {A13} (\bibinfo {year} {2016})},\ \Eprint
  {http://arxiv.org/abs/1502.01589} {arXiv:1502.01589 [astro-ph.CO]}
  \BibitemShut {NoStop}%
%%CITATION = ARXIV:1502.01589;%%
\bibitem [{\citenamefont {Aghanim}\ \emph {et~al.}(2018)\citenamefont {Aghanim}
  \emph {et~al.}}]{Aghanim:2018eyx}%
  \BibitemOpen
  \bibfield  {author} {\bibinfo {author} {\bibfnamefont {N.}~\bibnamefont
  {Aghanim}} \emph {et~al.} (\bibinfo {collaboration} {Planck}),\ }\href@noop
  {} {\  (\bibinfo {year} {2018})},\ \Eprint {http://arxiv.org/abs/1807.06209}
  {arXiv:1807.06209 [astro-ph.CO]} \BibitemShut {NoStop}%
%%CITATION = ARXIV:1807.06209;%%
\bibitem [{\citenamefont {{Beutler}}\ \emph {et~al.}(2011)\citenamefont
  {{Beutler}}, \citenamefont {{Blake}}, \citenamefont {{Colless}},
  \citenamefont {{Jones}}, \citenamefont {{Staveley-Smith}}, \citenamefont
  {{Campbell}}, \citenamefont {{Parker}}, \citenamefont {{Saunders}},\ and\
  \citenamefont {{Watson}}}]{Beutler2011}%
  \BibitemOpen
  \bibfield  {author} {\bibinfo {author} {\bibfnamefont {F.}~\bibnamefont
  {{Beutler}}}, \bibinfo {author} {\bibfnamefont {C.}~\bibnamefont {{Blake}}},
  \bibinfo {author} {\bibfnamefont {M.}~\bibnamefont {{Colless}}}, \bibinfo
  {author} {\bibfnamefont {D.~H.}\ \bibnamefont {{Jones}}}, \bibinfo {author}
  {\bibfnamefont {L.}~\bibnamefont {{Staveley-Smith}}}, \bibinfo {author}
  {\bibfnamefont {L.}~\bibnamefont {{Campbell}}}, \bibinfo {author}
  {\bibfnamefont {Q.}~\bibnamefont {{Parker}}}, \bibinfo {author}
  {\bibfnamefont {W.}~\bibnamefont {{Saunders}}}, \ and\ \bibinfo {author}
  {\bibfnamefont {F.}~\bibnamefont {{Watson}}},\ }\href {\doibase
  10.1111/j.1365-2966.2011.19250.x} {\bibfield  {journal} {\bibinfo  {journal}
  {mnras}\ }\textbf {\bibinfo {volume} {416}},\ \bibinfo {pages} {3017}
  (\bibinfo {year} {2011})},\ \Eprint {http://arxiv.org/abs/1106.3366}
  {arXiv:1106.3366} \BibitemShut {NoStop}%
\bibitem [{\citenamefont {{Ross}}\ \emph {et~al.}(2015)\citenamefont {{Ross}},
  \citenamefont {{Samushia}}, \citenamefont {{Howlett}}, \citenamefont
  {{Percival}}, \citenamefont {{Burden}},\ and\ \citenamefont
  {{Manera}}}]{Ross2015}%
  \BibitemOpen
  \bibfield  {author} {\bibinfo {author} {\bibfnamefont {A.~J.}\ \bibnamefont
  {{Ross}}}, \bibinfo {author} {\bibfnamefont {L.}~\bibnamefont {{Samushia}}},
  \bibinfo {author} {\bibfnamefont {C.}~\bibnamefont {{Howlett}}}, \bibinfo
  {author} {\bibfnamefont {W.~J.}\ \bibnamefont {{Percival}}}, \bibinfo
  {author} {\bibfnamefont {A.}~\bibnamefont {{Burden}}}, \ and\ \bibinfo
  {author} {\bibfnamefont {M.}~\bibnamefont {{Manera}}},\ }\href {\doibase
  10.1093/mnras/stv154} {\bibfield  {journal} {\bibinfo  {journal} {mnras}\
  }\textbf {\bibinfo {volume} {449}},\ \bibinfo {pages} {835} (\bibinfo {year}
  {2015})},\ \Eprint {http://arxiv.org/abs/1409.3242} {arXiv:1409.3242}
  \BibitemShut {NoStop}%
\bibitem [{\citenamefont {Alam}\ \emph {et~al.}(2017)\citenamefont {Alam} \emph
  {et~al.}}]{Alam:2016hwk}%
  \BibitemOpen
  \bibfield  {author} {\bibinfo {author} {\bibfnamefont {S.}~\bibnamefont
  {Alam}} \emph {et~al.} (\bibinfo {collaboration} {BOSS}),\ }\href {\doibase
  10.1093/mnras/stx721} {\bibfield  {journal} {\bibinfo  {journal} {Mon. Not.
  Roy. Astron. Soc.}\ }\textbf {\bibinfo {volume} {470}},\ \bibinfo {pages}
  {2617} (\bibinfo {year} {2017})},\ \Eprint {http://arxiv.org/abs/1607.03155}
  {arXiv:1607.03155 [astro-ph.CO]} \BibitemShut {NoStop}%
%%CITATION = ARXIV:1607.03155;%%
\bibitem [{\citenamefont {Betoule}\ \emph {et~al.}(2014)\citenamefont {Betoule}
  \emph {et~al.}}]{Betoule:2014frx}%
  \BibitemOpen
  \bibfield  {author} {\bibinfo {author} {\bibfnamefont {M.}~\bibnamefont
  {Betoule}} \emph {et~al.} (\bibinfo {collaboration} {SDSS}),\ }\href
  {\doibase 10.1051/0004-6361/201423413} {\bibfield  {journal} {\bibinfo
  {journal} {Astron. Astrophys.}\ }\textbf {\bibinfo {volume} {568}},\ \bibinfo
  {pages} {A22} (\bibinfo {year} {2014})},\ \Eprint
  {http://arxiv.org/abs/1401.4064} {arXiv:1401.4064 [astro-ph.CO]} \BibitemShut
  {NoStop}%
%%CITATION = ARXIV:1401.4064;%%
\bibitem [{\citenamefont {Asgari}\ \emph {et~al.}(2019)\citenamefont {Asgari}
  \emph {et~al.}}]{Asgari:2019fkq}%
  \BibitemOpen
  \bibfield  {author} {\bibinfo {author} {\bibfnamefont {M.}~\bibnamefont
  {Asgari}} \emph {et~al.},\ }\href@noop {} {\  (\bibinfo {year} {2019})},\
  \Eprint {http://arxiv.org/abs/1910.05336} {arXiv:1910.05336 [astro-ph.CO]}
  \BibitemShut {NoStop}%
%%CITATION = ARXIV:1910.05336;%%
\bibitem [{\citenamefont {de~Jong}\ \emph {et~al.}(2015)\citenamefont {de~Jong}
  \emph {et~al.}}]{deJong:2015wca}%
  \BibitemOpen
  \bibfield  {author} {\bibinfo {author} {\bibfnamefont {J.~T.~A.}\
  \bibnamefont {de~Jong}} \emph {et~al.},\ }\href {\doibase
  10.1051/0004-6361/201526601} {\bibfield  {journal} {\bibinfo  {journal}
  {Astron. Astrophys.}\ }\textbf {\bibinfo {volume} {582}},\ \bibinfo {pages}
  {A62} (\bibinfo {year} {2015})},\ \Eprint {http://arxiv.org/abs/1507.00742}
  {arXiv:1507.00742 [astro-ph.CO]} \BibitemShut {NoStop}%
%%CITATION = ARXIV:1507.00742;%%
\bibitem [{\citenamefont {Carter}\ \emph {et~al.}(2019)\citenamefont {Carter},
  \citenamefont {Beutler}, \citenamefont {Percival}, \citenamefont {DeRose},
  \citenamefont {Wechsler},\ and\ \citenamefont {Zhao}}]{Carter:2019ulk}%
  \BibitemOpen
  \bibfield  {author} {\bibinfo {author} {\bibfnamefont {P.}~\bibnamefont
  {Carter}}, \bibinfo {author} {\bibfnamefont {F.}~\bibnamefont {Beutler}},
  \bibinfo {author} {\bibfnamefont {W.~J.}\ \bibnamefont {Percival}}, \bibinfo
  {author} {\bibfnamefont {J.}~\bibnamefont {DeRose}}, \bibinfo {author}
  {\bibfnamefont {R.~H.}\ \bibnamefont {Wechsler}}, \ and\ \bibinfo {author}
  {\bibfnamefont {C.}~\bibnamefont {Zhao}},\ }\href@noop {} {\  (\bibinfo
  {year} {2019})},\ \Eprint {http://arxiv.org/abs/1906.03035} {arXiv:1906.03035
  [astro-ph.CO]} \BibitemShut {NoStop}%
%%CITATION = ARXIV:1906.03035;%%
\bibitem [{\citenamefont {Aker}\ \emph {et~al.}(2019)\citenamefont {Aker} \emph
  {et~al.}}]{Aker:2019uuj}%
  \BibitemOpen
  \bibfield  {author} {\bibinfo {author} {\bibfnamefont {M.}~\bibnamefont
  {Aker}} \emph {et~al.} (\bibinfo {collaboration} {KATRIN}),\ }\href {\doibase
  10.1103/PhysRevLett.123.221802} {\bibfield  {journal} {\bibinfo  {journal}
  {Phys. Rev. Lett.}\ }\textbf {\bibinfo {volume} {123}},\ \bibinfo {pages}
  {221802} (\bibinfo {year} {2019})},\ \Eprint
  {http://arxiv.org/abs/1909.06048} {arXiv:1909.06048 [hep-ex]} \BibitemShut
  {NoStop}%
%%CITATION = ARXIV:1909.06048;%%
\bibitem [{\citenamefont {Crittenden}\ and\ \citenamefont
  {Turok}(1996)}]{Crittenden:1995ak}%
  \BibitemOpen
  \bibfield  {author} {\bibinfo {author} {\bibfnamefont {R.~G.}\ \bibnamefont
  {Crittenden}}\ and\ \bibinfo {author} {\bibfnamefont {N.}~\bibnamefont
  {Turok}},\ }\href {\doibase 10.1103/PhysRevLett.76.575} {\bibfield  {journal}
  {\bibinfo  {journal} {Phys. Rev. Lett.}\ }\textbf {\bibinfo {volume} {76}},\
  \bibinfo {pages} {575} (\bibinfo {year} {1996})},\ \Eprint
  {http://arxiv.org/abs/astro-ph/9510072} {arXiv:astro-ph/9510072 [astro-ph]}
  \BibitemShut {NoStop}%
%%CITATION = ASTRO-PH/9510072;%%
\bibitem [{\citenamefont {Boughn}\ \emph {et~al.}(1998)\citenamefont {Boughn},
  \citenamefont {Crittenden},\ and\ \citenamefont {Turok}}]{Boughn:1997vs}%
  \BibitemOpen
  \bibfield  {author} {\bibinfo {author} {\bibfnamefont {S.~P.}\ \bibnamefont
  {Boughn}}, \bibinfo {author} {\bibfnamefont {R.~G.}\ \bibnamefont
  {Crittenden}}, \ and\ \bibinfo {author} {\bibfnamefont {N.~G.}\ \bibnamefont
  {Turok}},\ }\href {\doibase 10.1016/S1384-1076(98)00009-8} {\bibfield
  {journal} {\bibinfo  {journal} {New Astron.}\ }\textbf {\bibinfo {volume}
  {3}},\ \bibinfo {pages} {275} (\bibinfo {year} {1998})},\ \Eprint
  {http://arxiv.org/abs/astro-ph/9704043} {arXiv:astro-ph/9704043 [astro-ph]}
  \BibitemShut {NoStop}%
%%CITATION = ASTRO-PH/9704043;%%
\bibitem [{\citenamefont {Kimura}\ \emph {et~al.}(2012)\citenamefont {Kimura},
  \citenamefont {Kobayashi},\ and\ \citenamefont {Yamamoto}}]{Kimura:2011td}%
  \BibitemOpen
  \bibfield  {author} {\bibinfo {author} {\bibfnamefont {R.}~\bibnamefont
  {Kimura}}, \bibinfo {author} {\bibfnamefont {T.}~\bibnamefont {Kobayashi}}, \
  and\ \bibinfo {author} {\bibfnamefont {K.}~\bibnamefont {Yamamoto}},\ }\href
  {\doibase 10.1103/PhysRevD.85.123503} {\bibfield  {journal} {\bibinfo
  {journal} {Phys. Rev.}\ }\textbf {\bibinfo {volume} {D85}},\ \bibinfo {pages}
  {123503} (\bibinfo {year} {2012})},\ \Eprint {http://arxiv.org/abs/1110.3598}
  {arXiv:1110.3598 [astro-ph.CO]} \BibitemShut {NoStop}%
%%CITATION = ARXIV:1110.3598;%%
\bibitem [{\citenamefont {Spiegelhalter}\ \emph {et~al.}(2014)\citenamefont
  {Spiegelhalter}, \citenamefont {Best}, \citenamefont {Carlin},\ and\
  \citenamefont {van~der Linde}}]{RSSB:RSSB12062}%
  \BibitemOpen
  \bibfield  {author} {\bibinfo {author} {\bibfnamefont {D.~J.}\ \bibnamefont
  {Spiegelhalter}}, \bibinfo {author} {\bibfnamefont {N.~G.}\ \bibnamefont
  {Best}}, \bibinfo {author} {\bibfnamefont {B.~P.}\ \bibnamefont {Carlin}}, \
  and\ \bibinfo {author} {\bibfnamefont {A.}~\bibnamefont {van~der Linde}},\
  }\href {\doibase 10.1111/rssb.12062} {\bibfield  {journal} {\bibinfo
  {journal} {Journal of the Royal Statistical Society: Series B (Statistical
  Methodology)}\ }\textbf {\bibinfo {volume} {76}},\ \bibinfo {pages} {485}
  (\bibinfo {year} {2014})}\BibitemShut {NoStop}%
\bibitem [{\citenamefont {Creminelli}\ \emph {et~al.}(2019)\citenamefont
  {Creminelli}, \citenamefont {Tambalo}, \citenamefont {Vernizzi},\ and\
  \citenamefont {Yingcharoenrat}}]{Creminelli:2019kjy}%
  \BibitemOpen
  \bibfield  {author} {\bibinfo {author} {\bibfnamefont {P.}~\bibnamefont
  {Creminelli}}, \bibinfo {author} {\bibfnamefont {G.}~\bibnamefont {Tambalo}},
  \bibinfo {author} {\bibfnamefont {F.}~\bibnamefont {Vernizzi}}, \ and\
  \bibinfo {author} {\bibfnamefont {V.}~\bibnamefont {Yingcharoenrat}},\
  }\href@noop {} {\  (\bibinfo {year} {2019})},\ \Eprint
  {http://arxiv.org/abs/1910.14035} {arXiv:1910.14035 [gr-qc]} \BibitemShut
  {NoStop}%
%%CITATION = ARXIV:1910.14035;%%
\bibitem [{\citenamefont {de~Rham}\ and\ \citenamefont
  {Melville}(2018)}]{deRham:2018red}%
  \BibitemOpen
  \bibfield  {author} {\bibinfo {author} {\bibfnamefont {C.}~\bibnamefont
  {de~Rham}}\ and\ \bibinfo {author} {\bibfnamefont {S.}~\bibnamefont
  {Melville}},\ }\href {\doibase 10.1103/PhysRevLett.121.221101} {\bibfield
  {journal} {\bibinfo  {journal} {Phys. Rev. Lett.}\ }\textbf {\bibinfo
  {volume} {121}},\ \bibinfo {pages} {221101} (\bibinfo {year} {2018})},\
  \Eprint {http://arxiv.org/abs/1806.09417} {arXiv:1806.09417 [hep-th]}
  \BibitemShut {NoStop}%
%%CITATION = ARXIV:1806.09417;%%
\end{thebibliography}%

\end{document}